# Nonlinear Journal Prestige Normalized Productivity (JPNP): Defining and Measuring It

**Marek Kwiek**
Center for Public Policy Studies, Adam Mickiewicz University, Poznan, Poland
kwiekm@amu.edu.pl, ORCID: orcid.org/0000-0001-7953-1063

## Abstract

This article introduces Journal Prestige Normalized Productivity (JPNP), a nonlinear, percentile-normalized, and authorship-fractionalized indicator of research productivity. The indicator can be aggregated to the level of disciplines, institutions, and countries. Traditional measures based on simple publication counts and those not normalized by journal prestige do not seem to capture the current steep hierarchical structure of the global scientific publishing system. JPNP combines (a) percentile normalization of journal prestige, (b) a convex transformation that strengthens the upper tail of the journal distribution, and (c) authorship fractionalization. JPNP differentiates publication output according to the relative positions of journals and allocated fractions of contributions. Compared with linear weighting, JPNP increases the relative significance of publications in the highest journal percentiles and reduces the significance of publications in the lower journal percentiles. Percentile normalization provides a common scale of relative journal positions among fields of science. The comparability of the aggregated results also depends on conformity with the other assumptions of measurement. This article presents the construction of JPNP, its mathematical characteristics, and a sensitivity analysis of the selection of nonlinearity parameters. It also discusses the opportunities and limitations of the indicator in aggregated studies, institutional benchmarking, and science policy.

**Keywords:** research productivity; journal percentile ranks; academic journals; journal prestige; academic careers; inequality in science; academic publishing;

## 1. Introduction

Sociological and bibliometric studies of scientific knowledge and publication productivity have found extreme inequalities in the number of publications produced by individual researchers (Aguinis & O'Boyle, 2014; Kwiek, 2016; O'Boyle & Aguinis, 2012; Parker et al., 2010; Price, 1963; Serenko et al., 2011; Sidiropoulos et al., 2016). Early research on publication productivity treated individual productivity as a purely quantitative construct to measure the number of publications an individual researcher produces over a period of time (Cole & Cole, 1973; Crane, 1965). However, with the growth of research teams and different publication cultures among various disciplines, the approach of simple publication counting was no longer sufficient. Furthermore, because of the inequitable distribution of prestige among academic journals, reward systems within universities and national research assessment exercises do not view all journals as equal.

There are essentially four methods of measuring researchers' productivity. The first category utilizes basic publication counts. The second methodology incorporates co-authorship corrections, including fractional counting, harmonic counting, and similar variations. The third category includes productivity measures that utilize linear weighting factors for journal prestige. The final category includes methodologies that utilize percentile normalization and nonlinear weighting factors to account for the disproportionate concentration of prestige among high-prestige journals (Bak & Kim, 2019; Lindahl, 2018; Shibayama & Baba, 2015). Many fields have sets of elite journals that are recognized as being among the most prestigious journals in those respective fields (for economics, see Heckman & Moktan, 2020; for prestige economy of global higher education journals, see Kwiek, 2021a). Top researchers tend to publish in highly prestigious journals and less prestigious journals (Moed, 2005, p. 104).

Simple publication-based productivity metrics are easy to understand. Nevertheless, their limitations are well documented. For example, these metrics cannot differentiate between the quality of publications. These metrics also do not take into consideration how many co-authors contributed to each publication, and they rely heavily on discipline-specific norms and customs associated with publishing (Moed, 2005; Piro et al., 2013). Currently, they are primarily utilized as reference points (i.e., in sensitivity analyses) when making comparisons between different departments and/or institutions. However, they continue to be employed in comparative international academic profession studies that utilize survey-based data regarding published articles (Drennan et al., 2013).

Productivity measures were initially developed for individual-based research environments. Researchers working in highly collaborative environments often generate multiple papers through collaboration (Kwiek, 2021b). When each collaborator generates a paper and receives full credit for it, productivity counts inflate because of the contributions of all collaborators. Fractional methods and alternative credit allocation models were created to adjust for these inflated productivity counts (e.g., Egghe et al., 2000; Hagen, 2010; Stallings et al., 2013). Harmonic counting is defined as a methodology to improve the accuracy, fairness, and flexibility of standardized publication and citation credits (Hagen, 2010, p. 785). Although fractional methods provide better control of co-authorship inflation, they may punish researchers who work in highly collaborative fields, such as physics or astronomy, in which large numbers of researchers working together are considered normal (Larivière et al., 2015).

Another generation of productivity measures focused on the journal in which authors published their works rather than solely on the number of publications. Pinski and Narin (1971, p. 312) identified journal hierarchies that were constructed using citation structures and suggested developing “an influence weight of a journal,” in which some journals are “highly influential” and others are “peripheral” (see Kwiek, 2021a). A number of journal-specific output weights exist today, including the Journal Impact Factor (JIF), SCImago Journal Rank (SJR), Source Normalized Impact per Paper (SNIP), and prestige-weighted indicators based on citation networks (González-Pereira et al., 2010; Waltman & van Eck, 2013). González-Pereira et al. (2010) recommended a “journal prestige” indicator from the family of eigenvector-centrality indicators, and Guerrero-Bote and Moya-Anegón (2012) discussed the SJR2 indicator, in which citations are weighted by the prestige of the citing journal.

Visibility and potential influence are often linked to journal prestige (Bornmann, 2013). However, raw journal metrics are typically dependent on the field and do not lend themselves well to comparisons across fields, resulting in cross-field inequalities (Waltman & van Eck, 2013). The fourth generation of productivity measures uses the prestige of a journal, which normalizes prestige within a given field and then assigns nonlinear weights. The percentile normalization represents the relative position of a journal in a given field on a common scale. In the implementation presented, I use CiteScore Percentiles in the range of 0–99 (Bornmann, 2013; Leydesdorff & Bornmann, 2011). I distinguish between linear weighting, $y = x$, and nonlinear weighting, $y = x^k$, where $x$ belongs to [0,1] and $k > 1$. Linear prestige weighting is not identical to simple counting publications. In simple counting, each publication receives an equal prestige value of 1. Linear weighting is easy to understand: Two articles in a journal that is in the 40th journal percentile will weigh equally in calculating productivity as one article in a journal that is in the 80th percentile.

Linear weighting assumes that identical percentile growth means the same growth in weight. In JPNP, the same percentile growth more strongly differentiates weights in the upper part of the scale (Bornmann & Marx, 2014). Studies on the concentration of citations and inequalities in science generally justify interest in the upper segments of productivity distributions (Clauset et al., 2009; Kwiek & Roszka, 2024b; Radicchi et al., 2008). However, they do not indicate a specific function of weighting journal percentiles. In JPNP, the power function is an explicit construction assumption that allows for the regulation of the power to differentiate journals' positions. Heavy-tailed distributions reflect large rare events, as well as large fluctuations in the tail of the distribution (Clauset et al., 2008). In a given JPNP specification, the common exponent $k$ is applied. What changes between journals is the normalized percentile x rather than the exponent k. An increase of $k$ increases relative differences between the weights of publications from higher and lower segments of journals' hierarchy. These assumptions align with Pareto concentration and cumulative advantages (Allison, 1980; Allison & Stewart, 1974; Merton, 1968; Price, 1976), where the premium for publishing in high-prestige journals—very large in the journal-prestige distribution—grows disproportionately fast. In contrast, publications in lower parts of the distribution receive much less weight and attention from academic peers.

I propose JPNP as a nonlinear productivity measure based on the percentile normalization of journal prestige within scientific fields. The indicator involves the use of journal prestige percentiles (in the 0–99 range) as a cross-field comparable scale, their transformation to reflect the steep hierarchy of journals, and authorship fractionalization to reduce distortions caused by differences in team size across fields (Abramo et al., 2017). The best example in science, technology, engineering, and mathematics (STEM) fields is perhaps small-team mathematics versus large-team physics and astronomy. JPNP measures publication output by taking into account the relative positions of journals among other journals in a field and fractions of authorship. The position of a journal is the operationalization of publication channel prestige rather than a direct quality measure of particular articles. It is scalable, comparable across fields, and useful in current debates on inequality in science (Kwiek, 2018; Petersen et al., 2012; Wagner et al., 2015).

Contemporary research evaluation systems—national systems, such as the Polish research evaluations of 2022 and 2026, and international systems based on Scopus/SciVal and CiteScore Percentiles—generally assume that journals are not equal (CiteScore Percentile is a comprehensive metric used to measure the yearly citation impact of academic journals and other serial publications; SciVal brings together research performance, impact, and funding analytics within one connected environment. Both are developed by Elsevier and powered by Scopus data). An article published in the global top 1%–10% of journals has, on average, higher institutional and individual value than an article published in a low-prestige journal. For instance, in a recent Polish evaluation (2020–2025), which assessed disciplines within universities using the publication data of about 80,000 scientists, the difference in value between an article published in a low-prestige journal and one published in a high-prestige journal was forty-fold (5 vs. 200 points; the Polish points were based on a combination of the Scopus list with its percentiles, Web of Science list with its impact factors, and peer review provided by expert committees).

Simple publication counting allows for the identification of scientists with the highest volume of output without considering journal positions. Two scientists with ten publications in their publishing portfolios will have the same quantitative score, even if their publications have different prestige profiles. JPNP differentiates between such profiles because it responds to a different measurement question: How big is output considering assumed prestige weights (Aguinis & O'Boyle, 2014; Fox & Nikivincze, 2021; Kwiek, 2016, 2018; Parker et al., 2010; Serenko et al., 2011)? I focus on STEM fields, which are well represented in Scopus and Web of Science. These databases do not sufficiently cover the humanities and social sciences for the broad use of JPNP. I include only selected social science areas—Business, Management and Accounting; Economics, Econometrics and Finance; and Psychology—in which publishing cultures are closer to those of STEM and where journal articles play a central role in research output and academic careers (Kwiek & Roszka, 2024b).

JPNP does not require citations to accrue before articles are assessed. It may therefore serve to analyze publishing patterns in short time windows, also in early career stages. Ex ante interpretation refers to the position of a publication channel and requires the use of metrics available at the time of publication. This does not mean an independent assessment of the quality of a scientist's work. JPNP is more transparent methodologically than national journal scoring systems, which are usually based on a combination of peer reviews and bibliometrics. In practice, JPNP has been used to construct productivity deciles in 38 Organization for Economic Co-Operation and Development (OECD) countries (Kwiek & Szymula, 2025b), analyze mobility between productivity classes over time (Kwiek & Roszka, 2025), construct individual publishing and citation profiles and their institutional aggregates, and define the top 10% of scientists in terms of productivity (Kwiek & Roszka, 2024b) in order to define a scientific elite that combines high publication volume with publishing in high-prestige journals.

How strongly the logic of prestige-based stratification should be built into a productivity measure is a practical and theoretical decision. It can also be a policy decision for universities and entire science systems. Leading institutions may wish to value top-tier journals more strongly and devalue publications in lower-tier journals more than lower-tier institutions.

The article is of a conceptual and methodological nature. Its contribution is the formalization of the way of linking the percentile position of journals, nonlinear weights, and fractional authorship in an additive measure of research output. I present the characteristics of this construct and the implications of the selection of the $k$ parameter. Sensitivity analysis presented in Electronic Supplementary Material (ESM) serves the purpose of assessing the stability of results in a dataset studied rather than showing that JPNP is a universally more accurate measure of research quality.

The article is structured as follows: Section 2 presents Literature Review and Conceptual Framework; section 3 shows Operationalization and Computational Procedures; section 4 presents Discussion, and Section 5 – Conclusions and Implications. The article includes also extensive Electronic Supplementary Material.

## 2. Literature Review and Conceptual Framework

The rise of bibliometrics in evaluating research has led to a proliferation of author-level indicators of productivity. The lack of measurement tools was no longer an issue. Rather, the challenge was the limited ability to make comparable assessments. Wildgaard et al. (2014) demonstrated that 108 author-level indicators vary in structure, data needs, and computational requirements. Using clustering to assess 44 indicators, Wildgaard (2016) found that there is no universally accepted author-level productivity metric. Rather, the behavior of indicators varies depending on the discipline in which they are employed, the stage of a researcher's career, and the type of publication. Wildgaard (2016) noted that a gap exists between the output of a researcher reflected in their CV and the output represented by author-level indicators. Furthermore, the gap exists due to the undervaluation of certain achievements by author-level indicators. In describing the development and validity of 69 author-level indicators of research production, Wildgaard (2019) stated that "the author-level indicator market is a crowded marketplace" (p. 361).

Multiple counting methodologies and methods of allocating credits emerged. Gauffriau et al. (2008) identified three primary counting methodologies: whole-counting, fractional-counting, and first-author counting. Gauffriau (2021) identified 32 counting methodologies that had been developed since 1981 and assessed each methodology based on its mathematical characteristics and justifications. The findings were unambiguous: There is no neutral selection of counting methodologies. Each methodology produces different scores, changes the distribution of those scores, and affects the rankings of individuals and organizations (Gauffriau et al., 2007). The questions are as follows: Are there practical considerations supporting its utilization? How does it affect the scientific community? How do researchers perceive the indicator (Gauffriau, 2017)?

Bibliometrics also began to serve as an instrument for decision-making regarding academic careers and funding awards. González-Alcaide and Gorraiz (2018) provided an overview of the use of indicators for the assessment of researchers, as well as the need to match indicators to specific publishing norms within disciplines. Indicators eventually enter the governance of science (see also econometric approaches to measuring productivity in Daraio, 2019). Therefore, the choice of an indicator must depend on the goals of the study and its context.

Abramo and D'Angelo (2014) suggested developing a definition of productivity in an economic framework—as a function of output and input. In their view, simple counting of publications represents only a surrogate for actual productivity since it does not reflect either the quality of research or the amount of labor invested. Abramo and D'Angelo (2014) developed an indicator called fractional scientific strength (FSS), which combined four dimensions: publications, citations, fractional author contribution, and average annual salary. While the last dimension functions reasonably well when salaries are heavily influenced by academic rank—as is common in Italy and Poland—it functions poorly when salaries contain significant elements of competition (e.g., bonuses and individual contracts)—as in the U.S. and Chinese systems. FSS links the size of output and its citation impact with the information about inputs. JPNP assumes a different perspective; it measures output weighted by journal position, without considering workload or financing. The term "productivity" is thus used in a bibliometric sense rather than as a measure of economic effectiveness.

Research focused on the upper tail of distributions—productivity elites and top performers (Clauset et al., 2009)—plays a unique role in this area. Such research demonstrates that productivity is better understood by identifying distinct segments representing elite productivity rather than focusing solely on averages. Differences between these segments are larger, and mechanisms related to cumulative advantage are stronger (Agrawal et al., 2017; Fox & Nikivincze, 2021; Kwiek, 2018; Rosen, 1981). JPNP serves well a science system characterized by scarce prestige and resource concentration (Stephan, 2012; Whitley, 2000). Individuals and institutions behave similarly as prestige maximizers: Deriving prestige from high-status publications and grants enhances the likelihood of acquiring additional grants and further publishing in high-status outlets (Melguizo & Strober, 2007).

I consider academic careers to be shaped by the accumulation of successive advantages. Researchers collect publications, grants, honors, distinctions, awards, memberships in professional societies and academies, resources, equipment, collaborators, and networks. Successes during the early periods of one's career tend to generate subsequent successes and more access to resource and collaboration networks. Most advantages are generated via publishing—preferably numerous publications (volume) in high-quality journals (quality proxy). Data regarding 320,000 late-career scientists from 38 OECD countries demonstrate that more than half of the most productive scientists (scientists located in productivity Decile 10) were members of this group at earlier stages of their careers. Radical upward mobility across productivity deciles—Jumpers-Up (scientists moving between their early careers and mid-careers from Decile 1 to Decile 10)—occurred in about 0.5 percent of cases (0.51% in STEM, 0.56% in social sciences, and 0.51% in all disciplines combined; from 0.26% in Economics and Econometrics to 1.33% in Physics and Astronomy; Kwiek & Szymula, 2025b, p. 1024; see also Tay, 2026). Prior research has demonstrated a cumulative advantage in science (Allison, 1980; Allison & Stewart, 1974; Allison et al., 1982). Productivity distributions are strongly skewed and heavy-tailed (Price, 1976; Radicchi et al., 2008). High positions in a global journal hierarchy produce substantial returns and disproportionate benefits that disproportionately influence hiring and promotion processes in U.S.-based economics departments (Heckman & Moktan, 2020). The highest percentile tier for a journal in any area is reserved for those in the 99th percentile, such as *Science*, *Nature*, *Cell*, and *The Lancet*.

Any raw indicators of publication quality (e.g., JIF and SJR) vary across fields due to differing cultural norms regarding publishing and differing levels of citations per publication (Leydesdorff et al., 2019). Therefore, they require field-based normalization before being considered comparable.

JPNP allows the allocation of nonlinear weights to relative journal positions. The percentile determines the place in ordering, but it does not determine the size of prestige differences. The power transformation is a way to ascribe weights to these positions, and $k$ determines the power of this differentiation. Therefore, I introduced a convex transformation into JPNP:
$y = x^k$,
where $k > 1$ and $x \in [0,1]$.

In my implementation, $k = 2.5$.

To view the magnitude of weights applied and the growth rate at which premiums grow as journal prestige reaches increasingly higher percentiles, see Fig. 1.

Figure 1 clearly demonstrates that as you get closer to the top of journals' ranking, the marginal return for each additional contribution will be greater. As a result, small differences in journal prestige near the top produce a significantly larger difference in final scores than the same difference in prestige near the middle of the journal distribution. The shaded area of the top decile identifies where nonlinearity has its greatest impact on differentiating researchers. The figure shows the marginal weight increase resulting from the function ascribed rather than the assessment of career benefits from publishing. For $k > 1$, the percentile growth increases weight stronger in the upper part of the scale than in the lower part of the scale.

The scientist's JPNP in a given time window is the sum of the weighted contributions of their publications:

$$\mathrm{JPNP}_i(T) = \sum_{j=1}^{n_i(T)} c_{ij}\, x_j^k \qquad (1)$$

$c_{ij}$ means the contribution of the author $i$ in publication $j$, $x_j$ is the normalized percentile of the journal in which this publication was published, and $n_{i(T)}$is the number of publications of the author in time window $T$ (e.g., fractional or harmonic; see Hagen, 2010).

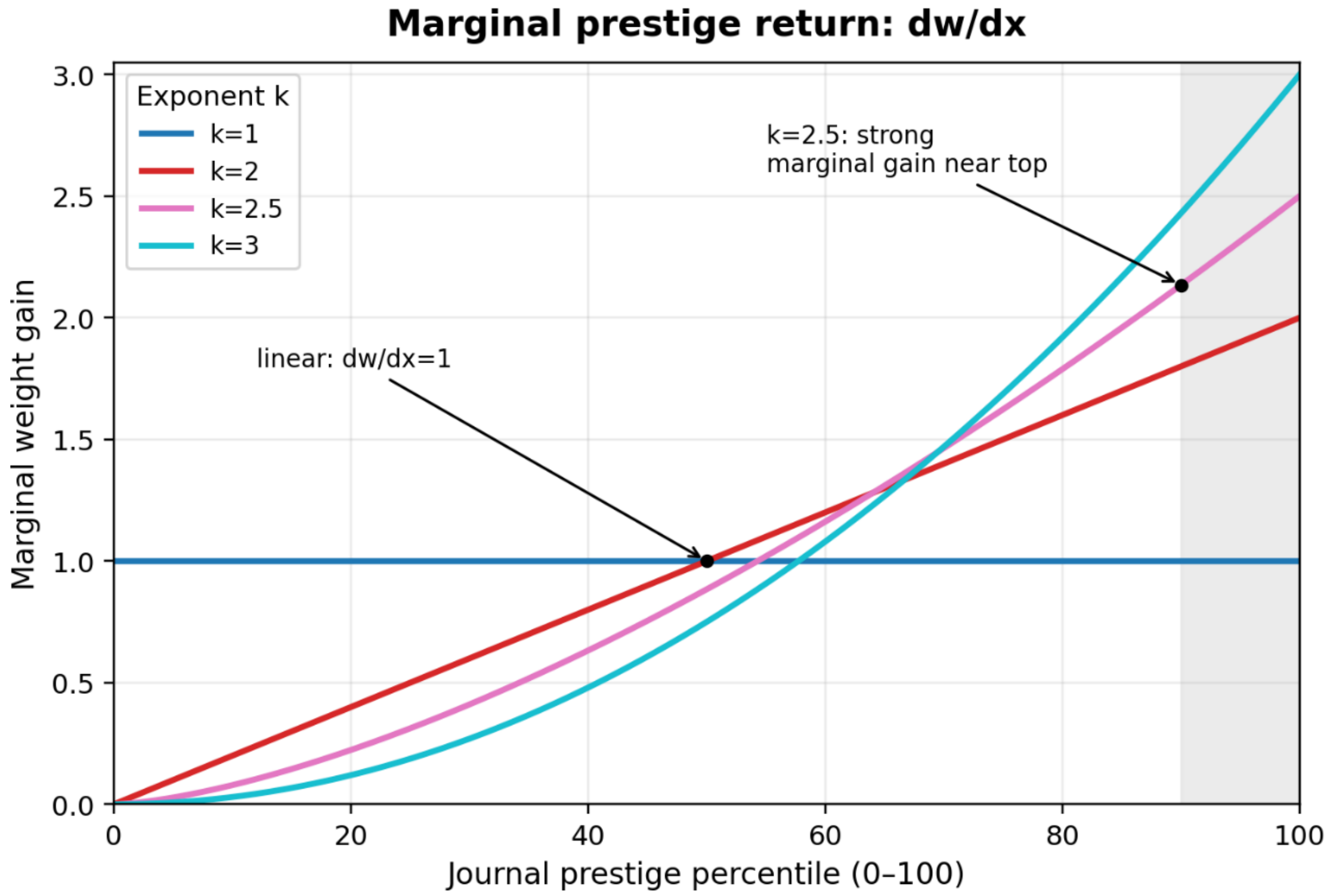

**Figure 1**. Marginal increase in prestige weight.

# 3. Operationalization and Computational Procedures

## 3.1. Data and Scope of Analysis

The calculations underlying previous analyses were based on Scopus raw data for Poland and for 38 OECD countries, provided by Elsevier's International Centre for the Study of Research (ICSR) Lab upon a multi-year agreement (see Kwiek & Roszka, 2024b, 2025; Kwiek & Szymula, 2025a, 2025b, 2026). The same procedures can be transferred to Web of Science or OpenAlex data, with some modifications and different scopes of data cleaning.

## 3.2. The Mathematical Core of JPNP: The Power Transformation of Prestige

Mathematically, JPNP is based on the power transformation $f(x) = x^{2.5}$, where $x \in [0,1]$ is the normalized journal prestige percentile. The function strictly increases since $f'(x) = 2.5x^{1.5} > 0$ for $x \in [0,1]$. It therefore preserves prestige ordering. It is also convex since

$$f''(x) = 3.75x^{0.5} > 0,$$

which means that it differentiates journals superlinearly in the upper part of the distribution and compresses its lower part. The transformation preserves the extremes, $f(0) = 0$ and $f(1) = 1$, which allows the linear and nonlinear versions to be compared on the same scale. For $x \in [0,1]$,

$$x^3 < x^{2.5} < x^2 < x.$$

Thus, the JPNP type is more nonlinear than the quadratic model, but it is less nonlinear than the cubic model. In comparison to the linear model, all other types have reduced weights, except for the extremes at 0 and 1. The function is continuous and mathematically stable across the interval [0,1].

## 3.3. Unit of Analysis, Publication Metadata, and Percentile Normalization

The primary unit of analysis in JPNP is the individual researcher. All of the researcher's publications from a specific period of time are used. These publications include both peer-reviewed journal articles and papers presented at conferences (Moed, 2005; Waltman, 2016). Metadata for each publication required for JPNP to create a publication profile are collected—journal name and CiteScore Percentile ranking, year published, number of co-authors, and author order—whenever possible. JPNP does not use citations of particular articles as a separate component. In the implementation presented, the journal position comes from CiteScore Percentile, that is, from a measure based on citation data at the journal level.

For each publication, CiteScore Percentile is used in a given field and year. Percentile $p_j$ from the scale 0–99 is rescaled to the interval [0,1] according to the formula $x_j = \frac{p_j}{99}$. It is a rescaling of the ready percentile rather than another determination of its position based on the number in the ranking.

$$x_j = 1 - \frac{r(J_j) - 1}{N_f - 1}, x_j \in [0,1]. \qquad (2)$$

In this scale, $x_j = 1$ corresponds to percentile 99, and $x_j = 0$ corresponds to percentile 0. Rescaling provides a common scale of relative journal positions across fields. However, it does not remove differences in publishing intensity, team size, or the structure of individual output. Comparability of journal positions, therefore, does not mean automatic comparability of aggregated productivity results. The upper 10% of journals does not cover the upper 10% of publications, as journals differ in the number of publications they publish. Thus, the normalization of journal positions does not automatically equalize the distribution of publishing opportunities or the share of publications per particular percentile. Figure 2 presents an intuitive illustration of why raw prestige metrics are not comparable across disciplines and why percentile normalization is needed.

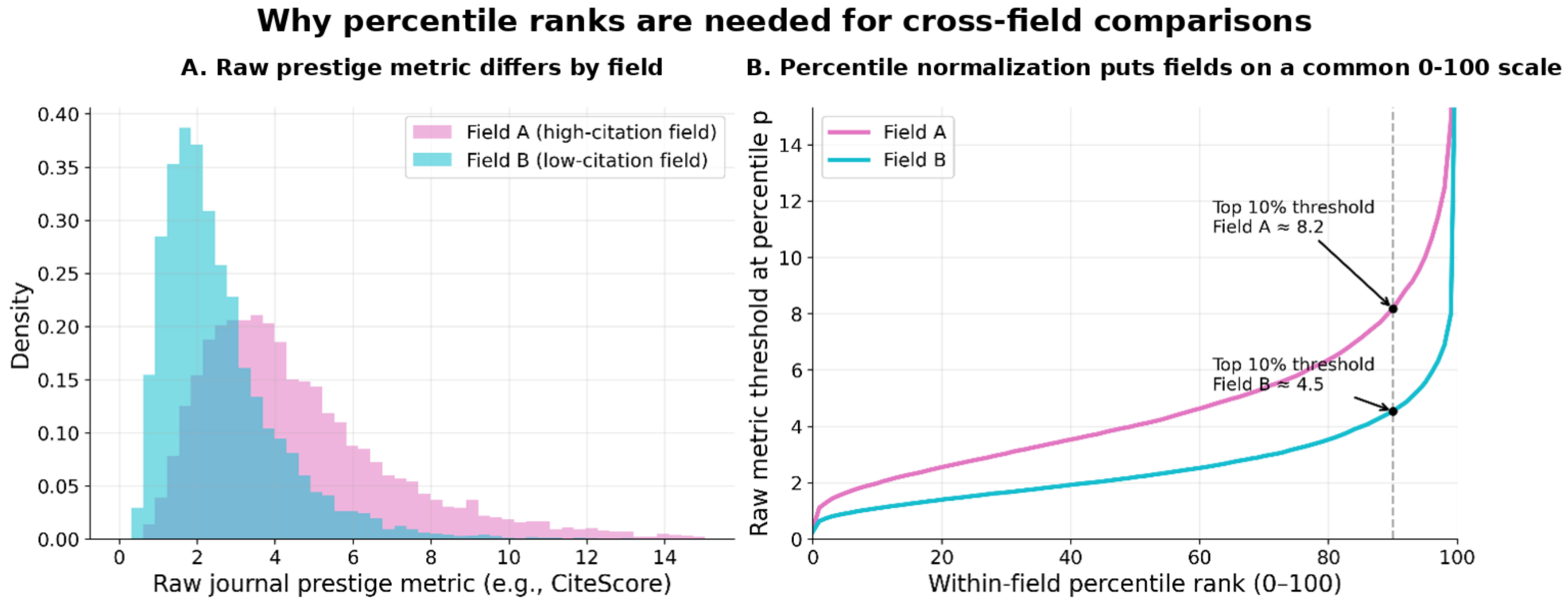

**Figure 2.** Why we need to use percentile ranks when making cross-disciplinary comparisons.

Panel A provides an illustration of this concept. It presents two stylized fields with different distributions of the same raw journal citability measure. Panel B shows that the same percentile position can correspond to different values of this measure. For example, if one of the disciplines has a high citation count, it generally has larger numbers, while another discipline with a low citation count has smaller numbers. Therefore, if we used an absolute raw score threshold to assess journal prestige, it would favor one discipline over the other. Panel B also demonstrates how the same ranked position (for instance, the 90th percentile or in the top 10%) represents different raw values in both disciplines. As such, percentiles allow us to compare our performance relative to our peers on a common relational basis regarding our positions in discipline-specific hierarchies.

## 3.4. Constructing Publication Weight: Prestige, Co-Authorship, and Aggregation into JPNP

The next step in constructing JPNP is to apply a nonlinear transformation of prestige:

$$w_j = x_j^k, \qquad (3)$$

where $k > 1$ controls sensitivity to the elite segment of journals. In the main analyses, $k = 2.5$ is used (Kwiek & Szymula, 2025b; see also Radicchi et al., 2008).

ESM shows measures of ranking stability among researchers via Spearman correlation coefficients, stability among top performers via Jaccard indices, entry and exit rates to the top 10% group, the effect of exponent $k$ on JPNP distributions, and relationships between JPNP and components of publication profiles. In the dataset examined and in the full counting variant of JPNP, $k = 2.5$ is in the range of values, guaranteeing high conformity of rankings and classifications.

A yearly Scopus data file that contains a field-specific percentile score for every journal indexed in Scopus can be utilized. The yearly Scopus data file may be accessed free of charge from the Scopus website (www.scopus.com). For the year 2026, this file contained 46,165 entries representing journals and conference proceedings.

Because co-authorship represents one of the most enduring aspects of scientific productivity (Wuchty et al., 2007), authorship weight is included in JPNP. Several other forms of publication counting are consistent with the structure of JPNP, including full counting, fractional counting $1/a_j$, positional harmonic weighting, and credit-share models (Hagen, 2010; Stallings et al., 2013). In the baseline fractional definition, I assume that $c_{ij} = \frac{1}{a_j}$, where $a_j$ means the number of all authors of publication j. The construct allows full counting in which $c_{ij} = 1$. The option used in sensitivity analysis in ESM is full counting (Egghe et al., 2000; Waltman, 2016).

In addition to correcting for co-authorship through the use of the authorship weight $c_{ij}$, Fig. 3 illustrates how various counting methodologies affect the amount of credit provided to a single author as the number of co-authors grows.

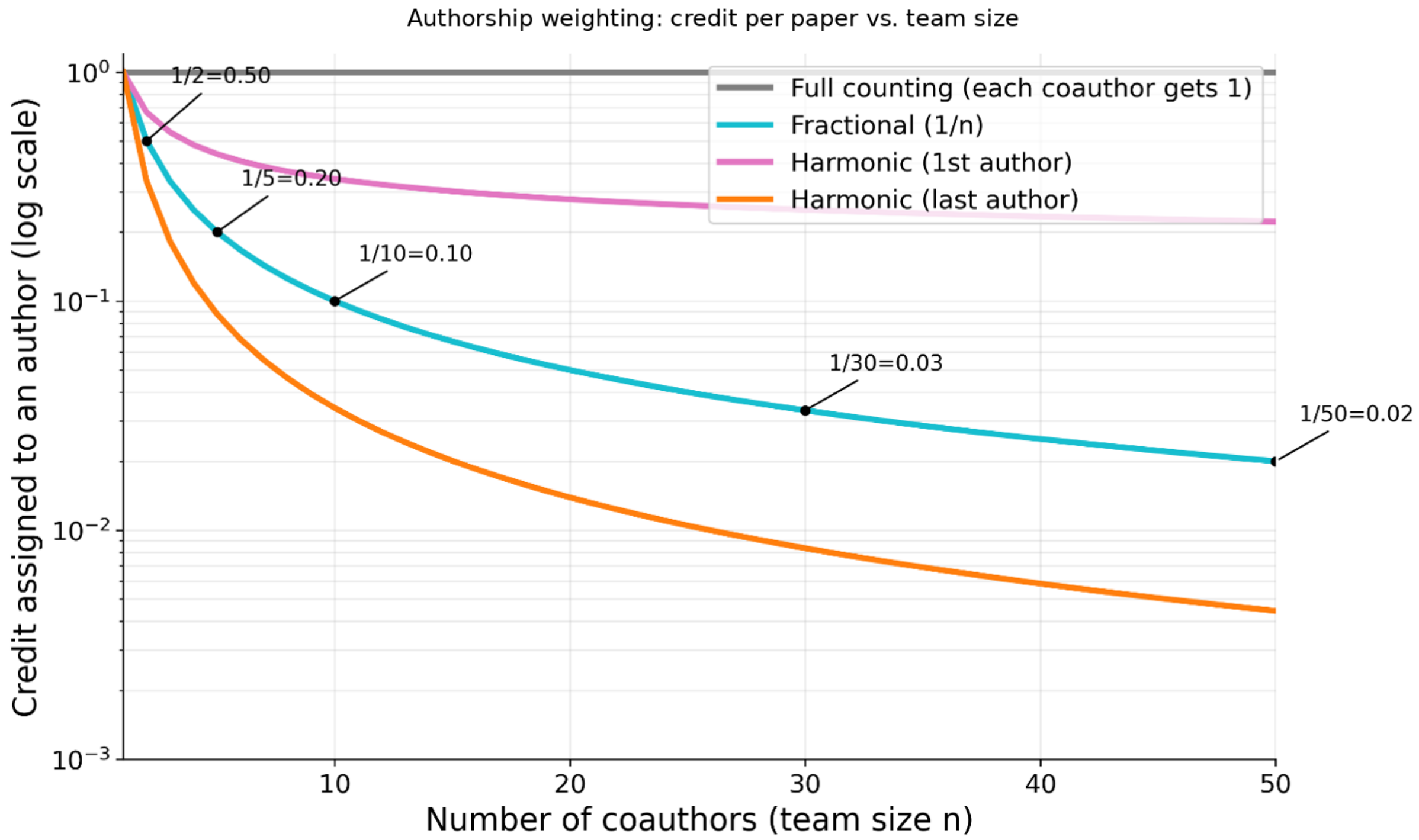

**Figure 3**. Authorship credit according to alternative counting strategies. Each point represents the fraction of credit attributed to a single author, depending on team size n.

Full counting assigns a value of 1 to each co-author, irrespective of the team size. Fields with many authors tend to receive inflated values when using full counting. Full counting ascribes the same publication as a whole to each co-author. Fractional counting limits the resulting increase in the sum of the author's credit but does not remove all cross-field differences. Fractional counting mechanically adjusts for this effect. As a result of decreasing by 1/n, multi-authored articles will contribute less than they would if counted fully. Harmonic counting also includes a position-dependent element. In the harmonic formula, the presented weights decrease as the position on the list of the authors grows. This is a rule of credit ascription rather than a measurement of real contributions by co-authors. Due to logarithmic scaling, these effects are clearly apparent, even for large teams, because the credit per author becomes negligible.

**Formal characteristics of JPNP.** For a given publication portfolio $\mathrm{JPNP_i} = \Sigma_j c_{ij} \times x_j^k$, where $c_{ij} \geq 0$, $x_j$ belongs to [0,1], and $k \geq 1$. From this definition, there are four basic characteristics. First, the result is limited: $0 \leq \mathrm{JPNP_i} \leq \Sigma_j c_{ij}$. The upper limit is, thus, the sum of the fractions of authorships. Second, the increase in the percentile of a single publication does not lower the score if the weight of authorship and the other components are constant. Third, the indicator is additive. The result for the whole output is the sum of the results of its mutually disjoint components, as long as the same percentiles, authorship weight, and *k* are retained. This allows JPNP to be decomposed into contributions of particular publications or groups of publications. Fourth, increasing *k* does not increase any individual results when percentiles and authorship weights are constant. However, it increases the relative advantage of higher percentile positions. For $0 < x_L < x_H \leq 1$, the proportion of their weights is $\left(\frac{x_H}{x_L}\right)^k$, and it increases as *k* increases. These characteristics result from the construction of the indicator rather than from an analysis of a dataset.

When authorship weights sum to 1, the joint contribution of all co-authors of a given publication is equal to the weight of this publication. The aggregation to institutions and countries additionally requires an explicit rule of ascribing affiliations and avoiding double counting. The sum of JPNP is dependent on the size of the unit analyzed and is not a measure of its effectiveness.

Calculations are done for a pre-defined time interval $T = [t_0, t_1]$. These time intervals often span a duration of around five or six years, such as 2018–2023. Increasing the length of time increases the stability of productivity classes and decreases random fluctuations from year to year (Kwiek & Roszka, 2025). Publications are allocated to the specified time interval based on the year of publication. The procedure consists of six steps: (1) collecting publications of scientist A in time window T, (2) ascribing CiteScore Percentiles to publications according to the rules of category and year, (3) rescaling percentiles $p_j$ to $x_j = \frac{p_j}{99}$, (4) calculating prestige weights $w_j = x_j^k$, (5) calculating authorship weights $c_{ij}$ following a given formula, and (6) summing contributions $c_{ij} \times w_j$. Finally, we can classify researchers into productivity classes based on their JPNP scores (for instance, top 1% or 10% in terms of productivity), applying a class-based approach that was used in the analysis of mobility between productivity classes over time (see Kwiek & Roszka, 2025) and in the study of scientific elite (Kwiek & Szymula, 2025b).

## 3.5. Interpreting Weighting Functions

The way we assign value through the weighting process is different from a nonlinear method for weighting (Table 1). Most significantly, there will be a stronger devaluation at the lower end of the journal ranking spectrum and a greater increase in the proportionate value placed on higher-ranked journals.

**Table 1**. Prestige weights for selected percentiles under four weighting schemes: linear ($y = x$) and nonlinear ($y = x^2, y = x^{2.5}, y = x^3$). Column $x$ (normalized) shows the journal percentile rank (0–99) rescaled to the interval $[0, 1]$. As the exponent $k$ increases, low and middle percentiles are increasingly depreciated, while the upper tail gains in relative importance.

| Percentile | $x$ (normalized) | $y = x$ | $y = x^2$ | $y = x^{2.5}$ | $y = x^3$ |
|---|---|---|---|---|---|
| 0 | 0.0000 | 0.0000 | 0.0000 | 0.0000 | 0.0000 |
| 10 | 0.1010 | 0.1010 | 0.0102 | 0.0032 | 0.0010 |
| 20 (bottom 20%) | 0.2020 | 0.2020 | 0.0408 | 0.0183 | 0.0082 |
| 30 | 0.3030 | 0.3030 | 0.0918 | 0.0505 | 0.0278 |
| 40 | 0.4040 | 0.4040 | 0.1632 | 0.1038 | 0.0660 |
| 50 (median) | 0.5051 | 0.5051 | 0.2551 | 0.1813 | 0.1288 |
| 60 | 0.6061 | 0.6061 | 0.3673 | 0.2860 | 0.2226 |
| 70 | 0.7071 | 0.7071 | 0.4999 | 0.4204 | 0.3535 |
| 80 | 0.8081 | 0.8081 | 0.6530 | 0.5870 | 0.5277 |
| 90 (top 10%) | 0.9091 | 0.9091 | 0.8264 | 0.7880 | 0.7513 |
| 95 (top 5%) | 0.9596 | 0.9596 | 0.9208 | 0.9020 | 0.8836 |
| 99 (top 1%) | 1.0000 | 1.0000 | 1.0000 | 1.0000 | 1.0000 |

The implications of selecting $k$ can be shown using two hypothetical single-authored publication portfolios. Portfolio A includes two articles in journals from the 50th percentile, while Portfolio B includes a single article in a journal from the 90th percentile. When $x$ = p/99 and $k$ = 1, the results are about 1.0101 and 0.9091, respectively, so the score is higher for Portfolio A. When $k$ = 2.5, the results are about 0.3625 and 0.7880, respectively, so the order is reversed. This example shows the implications of changing the weighting function rather than proving higher accuracy of either variant.

Figure 4 illustrates the differences among four different weighting models: $y = x$ (linear model), $y = x^2$ (quadratic model), $y = x^{2.5}$ (JPNP model), and $y = x^3$ (cubic model).

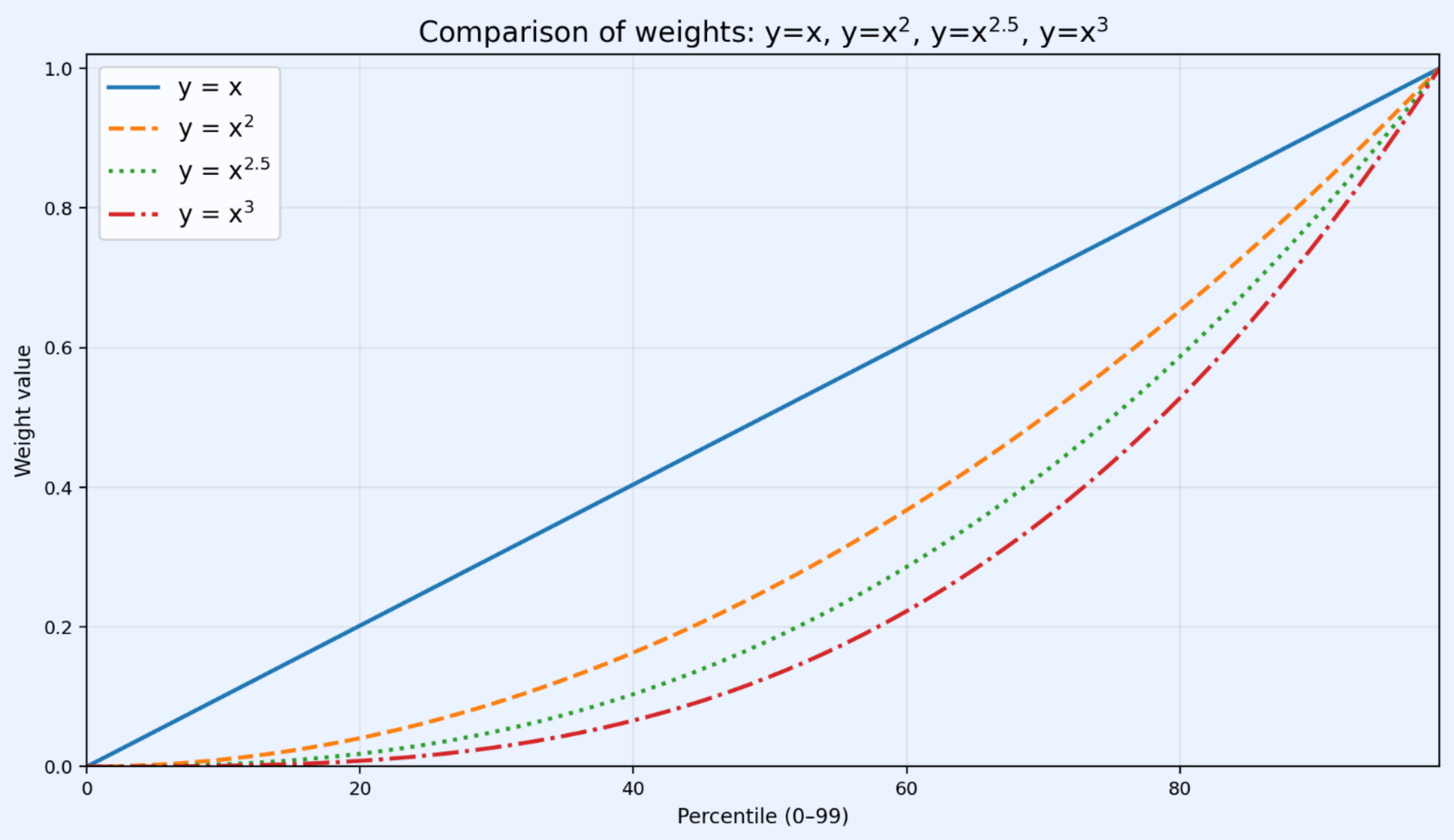

**Figure 4**. The differences among four different weighting models: $y = x$ (linear model), $y = x^2$ (quadratic model), $y = x^{2.5}$ (JPNP model), and $y = x^3$ (cubic model). As $k$ increases, the weight $x^k$ decreases for each value $0 < x < 1$. The proportional decrease is higher in lower journal percentiles. As a result, there is a relative increase in the number of higher positions because their absolute weights for $0 < x < 1$ are also increasing.

However, rather than illustrating absolute weights using a baseline formula, such as $y = x$, Figure 5 represents relative weights through changes in weight from a linear to a nonlinear approach. Each curve presents the difference $\Delta_k = x^k - x$, that is, nonlinear weight reduced by linear weight. For $k > 1$, the difference is negative within the interval $0 < x < 1$, and it equals 0 at both extremes.

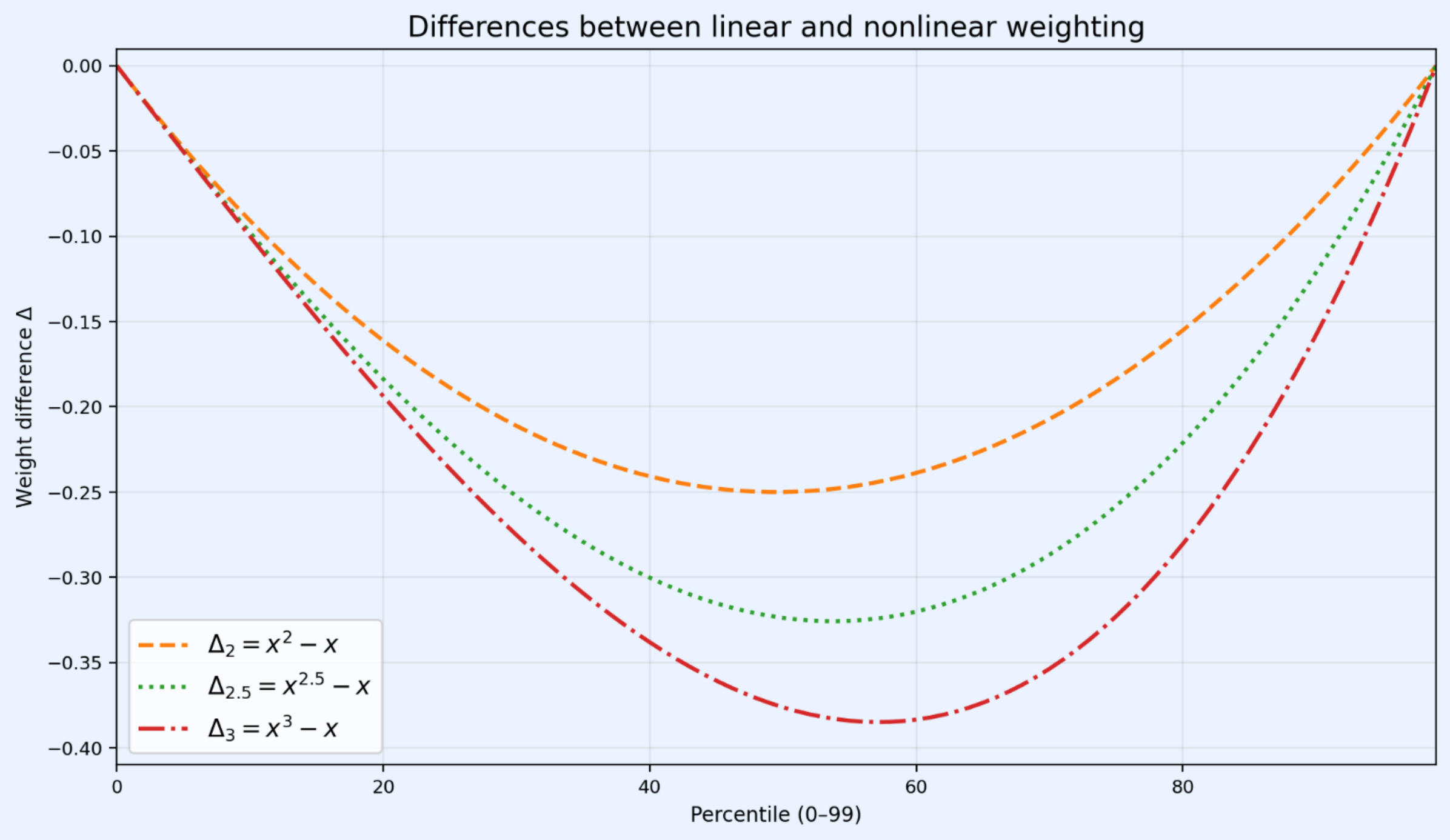

**Figure 5**. The difference between using a linear and nonlinear journal prestige weighting method. A curve for each $k$ = 2, 2.5 (default setting used in JPNP), and 3 displays the function $\Delta_k = x^k - x$, where x represents the normalized journal prestige percentile (0–99 converted into $[0, 1]$). A negative value means a lower weight ascribed to the same percentile position rather than a decline in journal prestige.

The minima show the highest absolute reduction of weight relative to the linear model. For $k$ = 2, 2.5, and 3, they occur when x is about 0.500, 0.543 and 0.577, respectively. After rescaling $x$ = p/99, it corresponds to percentiles 49.5, 53.7, and 57.2. The minimum differences are about −0.250, −0.326, and −0.385, respectively.

It should be noted that both extreme ends of the distribution (i.e., $x$ = 0 and $x$ = 1) result in zero differences (Δ = 0), regardless of whether we use a linear or nonlinear weighting method. Therefore, nonlinearity will add no additional weight to journals that comprise the absolute elite.

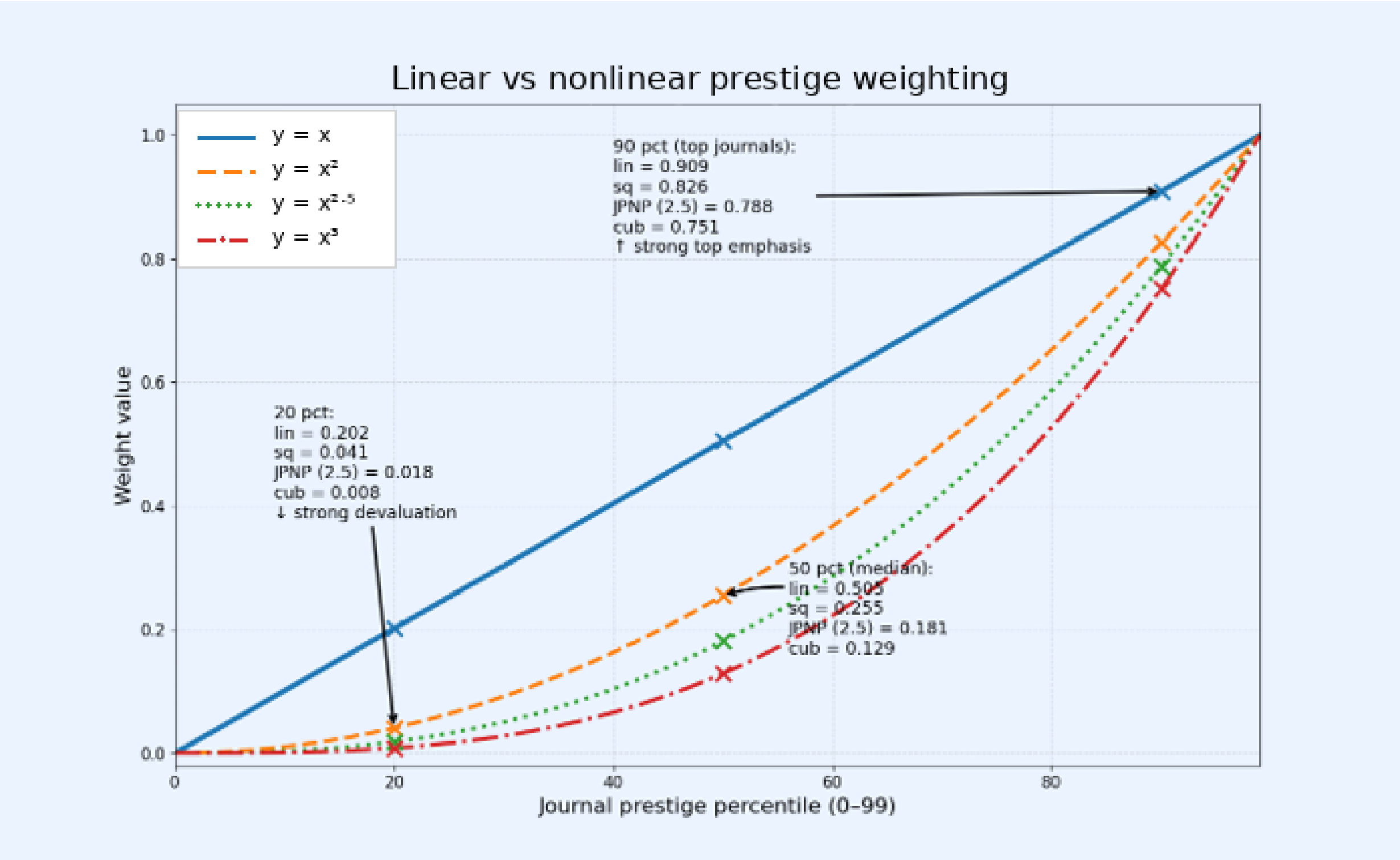

**Figure 6**. Journal prestige weights for linear ($y = x$) and nonlinear ($y = x^2$, $y = x^{2.5}$[JPNP], and $y = x^3$) models. These four models are compared using journal prestige percentiles rescaled to the interval [0,1].

At the 20th percentile, the linear model assigns a weight of 0.20. However, under nonlinear weighting, this decreases to approximately 0.04 when $k$ = 2, to about 0.018 when $k$ = 2.5, and to approximately 0.008 when $k$ = 3. Similarly, at the median point, the linear model yields a value of approximately 0.51. Using $x^2$ results in a value of about 0.26, $x^{2.5}$ in a value of about 0.18, and $x^3$ in a value of about 0.13.

Finally, at the 90th percentile point, high values of weight are present in both types of models, with approximate values of 0.91 in the linear model, 0.83 in $x^2$, 0.79 in $x^{2.5}$, and 0.75 in $x^3$. Therefore, the benefits to top journals come from the decreased values assigned to the bottom and middle portions of the distribution.

**Table 2**. Differences from the linear model: $\Delta_2 = x^2 - x$, $\Delta_{2.5} = x^{2.5} - x$, and $\Delta_3 = x^3 - x$. A negative sign shows how much nonlinear weighting reduces the weight of a specific percentile relative to $y = x$. The reductions are largest in the middle of the distribution and disappear in the top 1%, where all functions approach unity.

| Percentile | $\Delta_2 = x^2 - x$ | $\Delta_{2.5} = x^{2.5} - x$ | $\Delta_3 = x^3 - x$ |
|---|---:|---:|---:|
| 10 (bottom 10%) | −0.0908 | −0.0978 | −0.1000 |
| 20 | −0.1612 | −0.1837 | −0.1938 |
| 50 (median) | −0.2500 | −0.3238 | −0.3762 |
| 80 | −0.1551 | −0.2211 | −0.2804 |
| 90 (top 10%) | −0.0826 | −0.1211 | −0.1578 |
| 95 (top 5%) | −0.0388 | −0.0576 | −0.0760 |
| 99 (top 1%) | 0.0000 | 0.0000 | 0.0000 |

## 3.6. Calibration of the Parameter $k$, Publication Strategies, and the Properties of JPNP

JPNP is shown as a graphical representation of the strategy of a researcher's publication activity. Figure 7 shows a simplified map of the relationships between the fractionalized volume of publications and journals' average position. It illustrates the replaceability of both components in a model in which publications in a given publication portfolio have an identical percentile. In a real portfolio, JPNP also depends on the differentiation of percentiles; therefore, volume and a simple average journal prestige percentile do not unequivocally determine the result. Thus, an increased JPNP value may rise either by producing more papers, publishing these papers in more prestigious journals, or combining both strategies.

In a simplified model presented in Figure 7, various combinations of volume and average journal prestige percentile can provide the same value $\mathrm{FO} \times \mathrm{QP}^{\mathrm{k}}$. This makes possible the identification of four typical research profiles: 1. high-prestige hitters (low volume and high prestige): researchers with low quantities of papers published in highly prestigious journals; 2. volume producers (high volume and moderate prestige): researchers with many papers in moderately prestigious journals; 3. elite performers (high volume and high prestige): researchers with a high quantity of papers in the most prestigious journals; and 4. researchers with low volume and low average publication prestige.

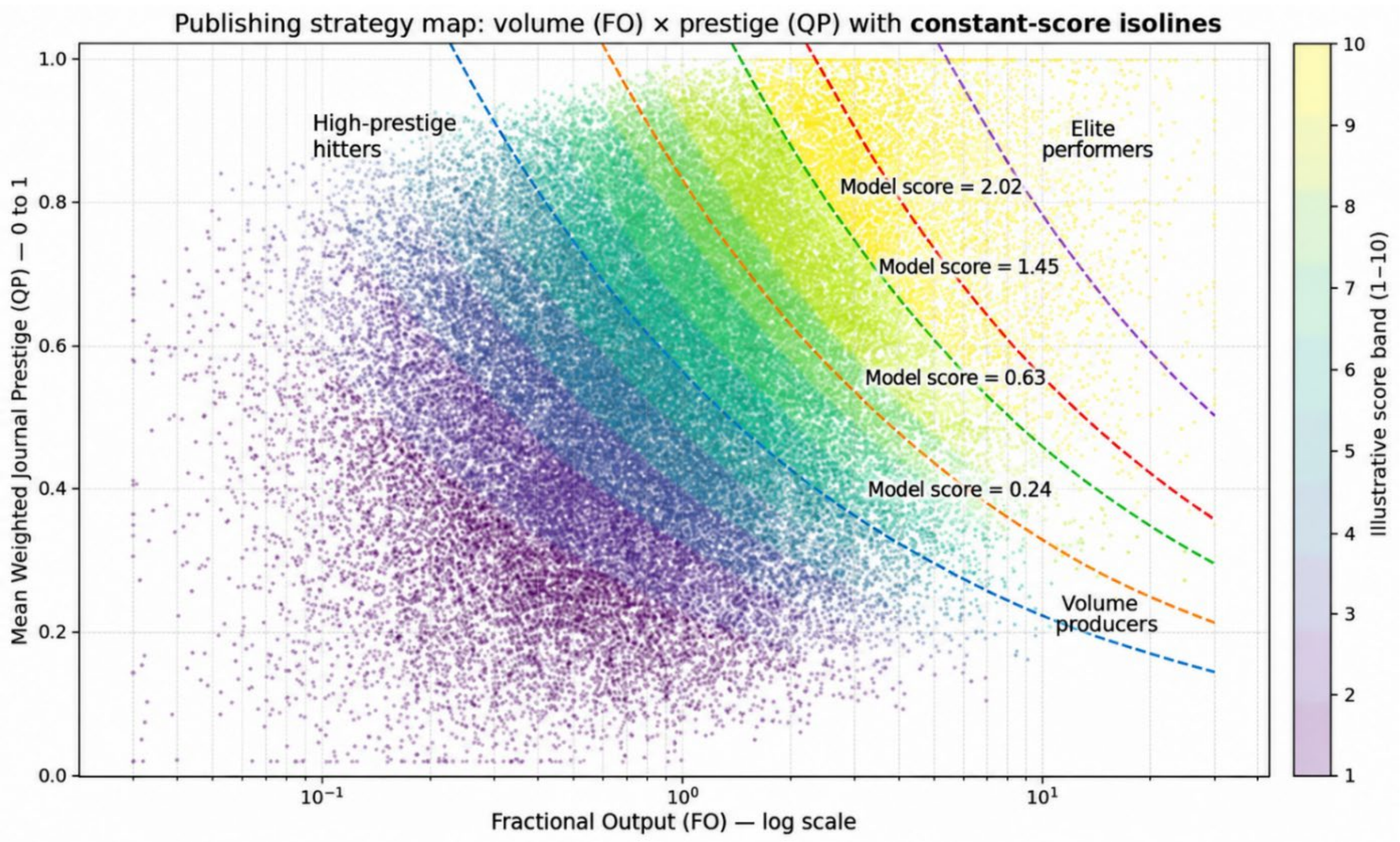

**Figure 7**. Simplified map of publishing strategies: volume × journal prestige (FO × QP). The x-axis presents the fractionalized volume of publications FO on a logarithmic scale. The y-axis presents an average weighted journal percentile QP in the range of [0,1]. Dashed lines link combinations providing the same value FO $\times$ $QP^k$. This value is equal to JPNP when publications in a given portfolio have an identical percentile. For portfolios with differentiated percentiles, this is a simplification. Colors refer to the relative bands of the model's values.
**Source:** Author's analytical illustration. No empirical data were used. The cloud of points presents the geometry of the simplified relationships between FO and QP. $k = 2.5$ is assumed.

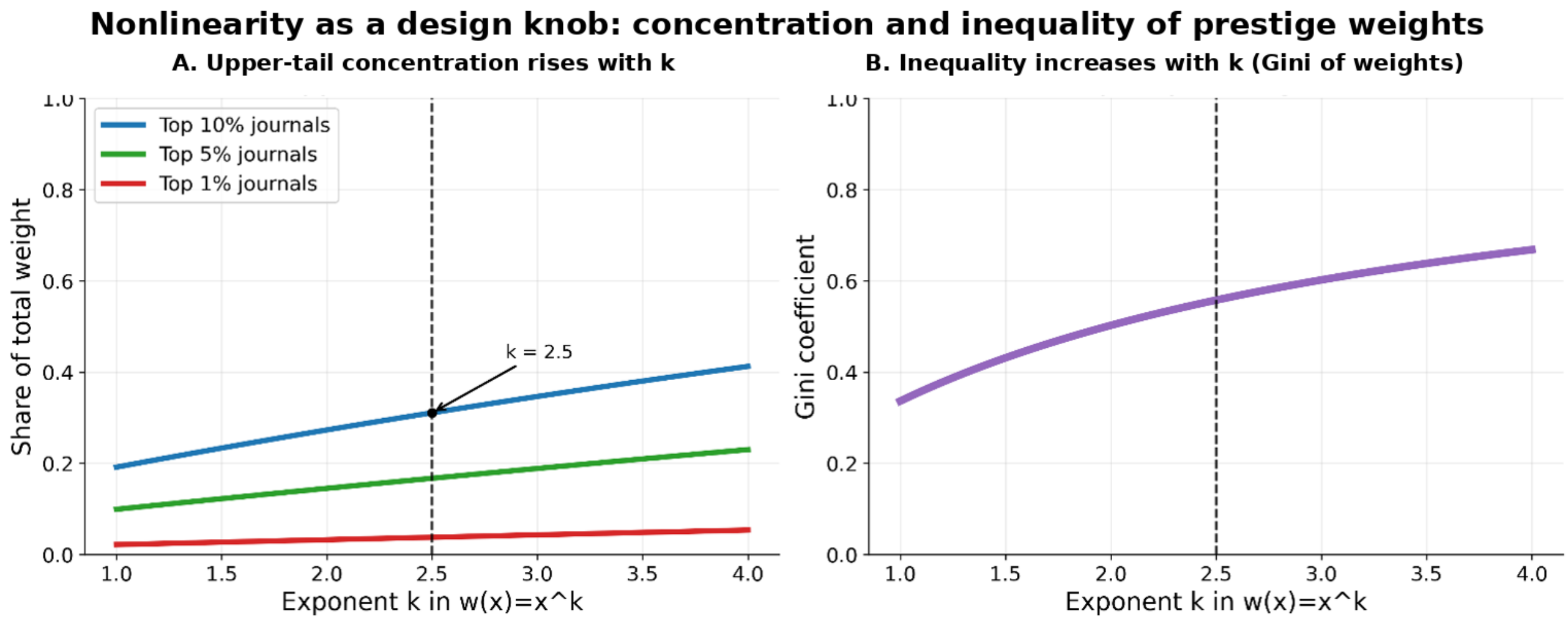

**Figure 8**. Nonlinearities as a calibration parameter: Concentration and inequality of prestige weights.

Panel A illustrates the portion of the overall weight assigned to the top journal segments (top 10%, top 5%, and top 1%) as a function of k. Panel B illustrates the Gini coefficient of the weight distribution. The dashed line represents $k = 2.5$. Concentration refers to weights in an assumed distribution of percentile references rather than the observed participation of top journals in the output of a selected population of scientists. The figure presents the consequences of a weighting function rather than an empirical distribution of JPNP.

Panel A shows how, as *k* increases, an increasing proportion of the total weight will reside at the upper end of the journal prestige hierarchy (i.e., in the top 10%, top 5%, and top 1%). Panel B shows how the Gini coefficient increases, with *k* indicating that the distribution of weights becomes increasingly skewed toward the elite. Therefore, the two panels can be viewed as a calibration chart illustrating what type of prestige policies are represented by JPNP. Additionally, these panels complement other figures by comparing only a limited number of discrete functions, showing how concentration increases monotonically with k.

In the dataset collected by the Observatory of Polish Science (Kwiek & Roszka, 2024a, 2024b) covering all Polish scientists who were identified via Scopus, JPNP exhibits significant right-skewness. Specifically, approximately 10% of all knowledge production in Poland occurs at the hands of the top 1% of Polish scientists working in 15 STEM disciplines and several social science areas. Approximately half of all knowledge production occurs at the hands of the top 10% (i.e., 10/50 rule; full counting JPNP; Kwiek & Roszka, 2024b). Both patterns are consistent with previous research concerning inequality in scientific productivity (Kwiek, 2016; Merton, 1968; Price, 1963; Seglen, 1992; Xie, 2014).

Nonlinear weighting changes the distribution of results, but its impact on dispersion and concentration depends on the structure of the publication portfolios. These characteristics need to be assessed for a given dataset. Even after omitting a subset of journals or researchers, percentiles can continue to divide the remaining sets or populations into the top 10% and bottom 10% categories.

Percentile normalization unifies the scale of journal positions, but in itself, it does not guarantee the similarity of distributions of JPNP or the reduction of differences between medians between the fields. The assessment of these characteristics requires a separate empirical comparison.

## 4. Discussion

At the discipline, institution, and country levels, JPNP highlights features of an author's publication portfolio that cannot be seen when considering either only productivity via publication count (linear) or without looking at the prestige of the journal where they were published. The use of JPNP therefore represents a shift in understanding what it means to measure productivity—moving from measuring productivity based solely on publication activity (i.e., volume) to measuring productivity as being related to journal prestige. Prestige can be defined and measured in multiple ways (e.g., González-Pereira et al., 2010; Guerrero-Bote & Moya-Anegón, 2012), using data from multiple sources (e.g., Scopus, Web of Science, and national journal reference lists), and can vary greatly depending on the source and the method used. Therefore, the underlying assumption of JPNP is very basic: All journals are not equal. There is a high level of stratification globally among journals, and scientists succeed in their careers by increasing measures based on publishing in the top tier of these hierarchies in their fields. The level of nonlinearity inherent in JPNP is represented by the exponent k, which can be calibrated over a wide range and is a way to tailor JPNP to meet the needs of various institutions and the requirements of national science systems at different stages of development.
However, direct comparisons require the application of the same *k* and consistent rules of normalization and ascription of authorship fractions. The results calculated using different values of *k* present different valuing systems and should not be treated as a single common scale.

In addition, it is important to note that most bibliometric indicators of publication productivity perform well at the level of larger units of aggregation (e.g., countries and institutions) rather than at the level of individual authors. I have applied JPNP primarily for intra-country comparisons (Poland) and inter-country comparisons (the 38 OECD countries and the 32 European countries) across 16 STEM and social sciences disciplines. My JPNP applications have aggregated individual authors' publications and citation portfolios into higher levels of aggregation. Individual results can be useful in studies of distributions, trajectories, and mobility between productivity classes. They should not guide the assessment of the research quality of a particular researcher or a decision about their employment, promotion, or financing.

JPNP remains a quantitative measure, but its results include the position of the publication channel. Prestige weights change the way output is aggregated; they do not replace a qualitative assessment of the contents of particular publications. Another possible proxy for quality would be the impact of those publications on global science (measured through citations). While classical linear measures of productivity are typically viewed as quantitative dimensions (Moed, 2005), JPNP moves productivity more toward value-weighted output metrics, similar to Abramo and D'Angelo's (2014) FSS.

JPNP illustrates a skewed productivity distribution. It uses an existing arrangement of journals, but the size of the differences between the results also depends on the weighting function assumed. The indicator is therefore not a neutral reconstruction of inequalities; it operationalizes them according to an explicit choice of parameter k. Publications from middle percentiles keep their positive contribution, but their weights are lower than in a linear model, and their relative significance to the highest percentiles decreases. When comparing journals across fields with different publication cultures (cross-field comparisons), percentile normalizations are crucial. Unlike some citation indicators that require complex normalization procedures (Waltman & van Eck, 2013), journal percentiles provide relatively stable comparisons.

There are, however, several limitations associated with JPNP. Most of all, journal prestige rankings do not account for variations in the quality of individual articles. While I have argued for $k = 2.5$ as a reasonable general case, $k$ may need additional calibration depending on the field and country in question. Future studies will develop composite indicators combining JPNP with citations (I call this combination JPNP 2.0), apply dynamic versions of productivity to different types of career paths, and test alternative forms of nonlinearity for different prestige regimes—including national journal lists employed by national research evaluation agencies (as in Poland and Norway).

JPNP introduces new methods of conceptualizing and quantifying productivity. It provides a tool consistent with the current stratification of global science—identifying elites in publication activity, elite institutions, and publication patterns across disciplines, institutions, and countries. Additionally, JPNP demonstrates structural inequity in productivity—examples include Poland (Kwiek & Roszka, 2024b) and OECD countries (Kwiek & Szymula, 2025b). JPNP reduces the contribution of publications from journals with lower percentiles according to the weighting function adopted. This does not mean that these publications are devoid of scientific value.

## 5. Conclusions and Implications

The fundamental limitation of JPNP is the ascription of weight to publications based on journal positions rather than an assessment of their content. In the implementation presented in this paper, the journal position is determined by CiteScore Percentile. The ex ante interpretation is justified when the metric available at the moment of publication is used. However, JPNP is not a direct measure of article quality or its future citation impact. Peer-reviewed methods of differentiating among journals, although viewed often as objective, are, in fact, highly subjective and not feasible for large-scale studies. At the journal level, journal prestige indicates the expected average visibility of articles published in the journal, as well as the expected average impact of those articles expressed in citations. However, journal prestige does not indicate variability in the impact of articles at the article level. Articles with high impact are often found in non-top-tier journals, while articles with low impact are often found in top-tier journals. This mismatch represents one of the largest limitations of journal prestige when evaluating individual performance; therefore, peer review remains essential in determining whether an individual has achieved success through their work. JPNP will be less problematic at a higher level of aggregation (i.e., examining entire institutions or countries).

CiteScore Percentile is the quotient of citations received in a given four-year window by the relevant types of documents published in the same window and the number of these documents. CiteScore Percentile then describes the respective positions of journal titles in a field category on the basis of CiteScore. When there is a small number of documents, single highly cited articles can powerfully inflate the average value.

Journal prestige measures are not static across bibliometric databases and can vary slightly over time. Additionally, the number of indexed journals is increasing. This will affect the ability of each journal percentile to accurately reflect the distribution of journals within a subfield—as there is an increase in the number of journals, there is also an increase in the number of journals within each percentile, even in the case of top percentiles (for example, the number of top 1% journals within a subfield). Normalizing percentiles within a field will help reduce the scale issue related to differences in numbers of journals across various fields. However, journal prestige is still dynamic, particularly in fast-moving fields.

Although Scopus has tracked 46,165 journals and conference proceedings (2026), percentile positions can vary each year. Thus, journal rank studies require additional robustness tests: other metrics (SJR or SNIP), temporal analysis of journal rank changes over the years, and comparison of rankings based on different field classifications. The assignment of an author's paper into a specific discipline within the citation base may also result in a shift in its percentile position.

The fractionalization of authorship in calculating productivity represents an increasing trend toward "team science" in academia; however, it fails to account for the variability in individual co-authors' contributions to papers both within and between disciplines. More complex rules of credit ascription (e.g., harmonic models; see Hagen, 2010) may better present particular authorship practices, but they do not guarantee higher accuracy. Their application requires a justification of assumptions made and the conducting of sensitivity analyses.

Therefore, the selection of the fractionalization method should be supported by empirical evidence and evaluated through sensitivity analysis. JPNP ascribes weights determined by the function adopted to relative journal positions. The adequacy of this function needs to be considered with respect to the aim of analysis and principles of valuing output rather than as an automatic reflection of a single real prestige scale. The value selected for the exponent $k$ in the nonlinear function may need to be calibrated relative to discipline and the changing structure of prestige among journals in a particular system, country, journal point system, journal listing, and publication cultures. Therefore, JPNP's nonlinearity parameter should be considered a tunable parameter, particularly when comparing fields and when analyzing long-term trends.

In some fields and science systems, the nonlinearity parameter may have to be set higher than is suggested here. Higher *k* may be welcomed by institutional policy that gives a stronger advantage to publishing in highly prestigious journals. Such a choice determines the way in which scientific output is valued in a given system. This does not mean, though, that publications in lower journal percentiles do not contribute to the development of science.

JPNP covers only a particular dimension of publication output (journal prestige). It does not consider many other results of scientific work, such as patenting, software, and databases, nor wider dimensions of academic work, such as teaching, mentoring, organizational work, grants, social impact, and open science (Todeschini & Baccini, 2016, p. 335). Similar to all publication-related metrics, JPNP is best suited for providing insight into publication productivity at the aggregate level.

From the perspective of science policy, JPNP allows for the comparison of outputs using the explicitly shown principles of prestige weighting. Percentile normalization provides a common scale of journal positions, but it does not remove all differences among disciplines. Applications in national research evaluation and institutional benchmarking require consistent observation windows, authorship rules, and *k* value, as well as the consideration of the size of units under comparison. On the other hand, JPNP clearly distinguishes the upper portion of the journal distribution, making it easier to identify productive scientists and support the development of programs to encourage excellence in research and funding competitions.

JPNP reduces the relative value of increasing the volume of publications in journals with lower percentiles. However, it still gives an advantage to the number of publications with the same prestige value and authorship. Thus, the indicator does not include a mechanism that defends against dividing the results into many articles (salami slicing); its behavioral implications require a separate study. The results describe output measured by journal position rather than independently assessed research quality.

I also use a hybrid model (JPNP 2.0), which links journal prestige to field-normalized citations—specifically to four-year field-weighted citation impact values per publication (not studied in this paper). JPNP enables studies of entry into high-productivity classes, persistence in the publication elite, and departure from it. Therefore, JPNP enables me to test whether individual productivity is stable or unstable over time (Kwiek & Roszka, 2025; Kwiek & Szymula, 2025b).

Comparative future uses of JPNP in a global context could have substantial value. The analyses may identify elite groups of researchers who publish in highly prestigious journals and allow for the study of the relationship between past and current JPNP levels and continued publishing throughout later career stages.

The normative implications of using JPNP as a measure of scientific productivity are also very important. There simply is not enough space available in highly prestigious journals to allow science policy for large numbers of institutions to require massive proportions of their faculty to publish in those journals. Therefore, while journal prestige-sensitive measures may help identify scientists who conduct prestigious research, over time, these measures may exacerbate existing inequalities and create strategic behaviors among scientists and institutions. Thus, ethical research is needed regarding the long-term negative unintended effects of nonlinear productivity measures.

## Acknowledgments

I gratefully acknowledge the support of Dr. Wojciech Roszka, particularly in providing electronic supplementary material.

## Competing interests

The author has no competing interests.

## Funding information

This publication was co-financed from the state budget as part of the Minister of Education and Science's program "Science for Society II," project No. NdS-II/SP/0010/2023/01, with a grant amount of 750,000 PLN and a total project value of 750,000 PLN.

## Data availability

I used data from Scopus, a proprietary scientometric database. For legal reasons, data from Scopus received through collaboration with the ICSR Lab (Elsevier) cannot be made openly available.

## References

Abramo, G., & D'Angelo, C. A. (2014). How do you define and measure research productivity? *Scientometrics*, *101*(2), 1129–1144.

Abramo, G., D'Angelo, C. A., & Soldatenkova, A. (2017). An investigation on the skewness patterns and fractal nature of research productivity distributions at field and discipline level. *Journal of Informetrics*, *11*(1), 324–335.

Agrawal, A., McHale, J., & Oettl, A. (2017). How stars matter: Recruiting and peer effects in evolutionary biology. *Research Policy*, *46*(4), 853–867.

Aguinis, H., & O'Boyle, E. (2014). Star performers in twenty-first century organizations. *Personnel Psychology*, *67*(2), 313–350.

Allison, P. D. (1980). Inequality and scientific productivity. *Social Studies of Science*, *10*, 163–179.

Allison, P. D., & Stewart, J. A. (1974). Productivity differences among scientists: Evidence for accumulative advantage. *American Sociological Review*, *39*(4), 596–606.

Allison, P. D., Long, J. S., & Krauze, T. K. (1982). Cumulative advantage and inequality in science. *American Sociological Review*, *47*(5), 615–625.

Bak, H. J., & Kim, D. H. (2019). The unintended consequences of performance-based incentives on inequality in scientists' research performance. *Science and Public Policy*, *46*(2), 219–231.

Bornmann, L. (2013). What is societal impact of research and how can it be assessed? A literature survey. *Journal of the Association for Information Science and Technology*, *64*(2), 217–233.

Bornmann, L., & Marx, W. (2014). How to evaluate individual researchers working in the natural and life sciences meaningfully? A proposal of methods based on percentiles of citations. *Scientometrics*, *98*, 487–509.

Clauset, A., Shalizi, C. R., & Newman, M. E. J. (2009). Power-law distributions in empirical data. *SIAM Review*, *51*(4), 661–703. https://www.doi.org/10.1137/070710111.

Cole, J. R., & Cole, S. (1973). *Social stratification in science*. University of Chicago Press.

Crane, D. (1965). Scientists at major and minor universities: A study of productivity and recognition. *American Sociological Review*, *30*(5), 699–714.

Daraio, C. (2019). Econometric approaches to the measurement of research productivity. In W. Glänzel, H. F. Moed, U. Schmoch, & M. Thelwall (Eds.), *Springer handbook of science and technology indicators* (pp. 633–666). Springer.

Drennan, J., Clarke, M., Hyde, A., & Politis, Y. (2013). The research function of the academic profession in Europe. In U. Teichler & E. A. Höhle (Eds.), *The work situation of the academic profession in Europe: Findings of a survey in twelve countries* (pp. 109–136). Springer.

Egghe, L., Rousseau, R., & van Hooydonk, G. (2000). Methods for accrediting publications to authors or countries: Consequences for evaluation studies. *Journal of the American Society for Information Science*, *51*(2), 145–157.

Fox, M. F., & Nikivincze, I. (2021). Being highly prolific in academic science: Characteristics of individuals and their departments. *Higher Education*, *81*, 1237–1255.

Gauffriau, M. (2017). A categorization of arguments for counting methods for publication and citation indicators. *Journal of Informetrics*, *11*(3), 672–684.

Gauffriau, M. (2021). Counting methods introduced into the bibliometric research literature 1970–2018: A review. *Quantitative Science Studies*, *2*(3), 932–975.

Gauffriau, M., Larsen, P. O., Maye, I., Roulin-Perriard, A., & von Ins, M. (2007). Publication, cooperation and productivity measures in scientific research. *Scientometrics*, *73*(2), 175–214.

Gauffriau, M., Larsen, P. O., Maye, I., Roulin-Perriard, A., & von Ins, M. (2008). Comparisons of results of publication counting using different methods. *Scientometrics*, *77*(1), 147–176.

González Alcaide, G., & Gorraiz, J. I. (2018). Assessment of researchers through bibliometric indicators: The area of information and library science in Spain as a case study (2001–2015). *Frontiers in Research Metrics and Analytics*, *3*, Article 15.

González-Pereira, B., Guerrero-Bote, V. P., & Moya-Anegón, F. (2010). A new approach to the metric of journals' scientific prestige: The SJR indicator. *Journal of Informetrics*, *4*(3), 379–391.

Guerrero-Bote, V. P., & Moya-Anegón, F. (2012). A further step forward in measuring journals' scientific prestige: The SJR2 indicator. *Journal of Informetrics*, *6*(4), 674–688.

Hagen, N. T. (2010). Harmonic publication and citation counting: Sharing authorship credit equitably - Not equally, geometrically or arithmetically. *Scientometrics*, *84*(3), 785–793.

Heckman, J. J., & Moktan, S. (2020). Publishing and promotion in economics: The tyranny of the top five. *Journal of Economic Literature*, *58*(2), 419–470.

Kwiek, M. (2016). The European research elite: A cross-national study of highly productive academics in 11 countries. *Higher Education*, *71*(3), 379–397. https://doi.org/10.1007/s10734-015-9910-x

Kwiek, M. (2018). High research productivity in vertically undifferentiated higher education systems: Who are the top performers? *Scientometrics*, *115*(1), 415–462. https://doi.org/10.1007/s11192-018-2644-7

Kwiek, M. (2021a). The prestige economy of higher education journals: A quantitative approach. *Higher Education*, *81*, 493–519. https://doi.org/10.1007/s10734-020-00553-y

Kwiek, M. (2021b). What large-scale publication and citation data tell us about international research collaboration in Europe: Changing national patterns in global contexts. *Studies in Higher Education*, *46*(12), 2629–2649. https://doi.org/10.1080/03075079.2020.1749254

Kwiek, M., & Roszka, W. (2024a). Once highly productive, forever highly productive? Full professors' research productivity from a longitudinal perspective. *Higher Education*, *87*, 519–549 (2024). https://doi.org/10.1007/s10734-023-01022-y

Kwiek, M., & Roszka, W. (2024b). Top research performance in Poland over three decades: A multidimensional micro-data approach. *Journal of Informetrics*, *18*(4), 101595. https://doi.org/10.1016/j.joi.2024.101595

Kwiek, M., & Roszka, W. (2025). Are scientists changing their research productivity classes when they move up the academic ladder? *Innovative Higher Education*, *50*, 329–367. https://doi.org/10.1007/s10755-024-09735-3

Kwiek, M., & Szymula, L. (2025a). Quantifying attrition in science: A cohort-based, longitudinal study of scientists in 38 OECD countries. *Higher Education*, *89*, 1465–1493. https://doi.org/10.1007/s10734-024-01284-0

Kwiek, M., & Szymula, Ł. (2025b). Quantifying lifetime productivity changes: A longitudinal study of 320,000 late-career scientists. *Quantitative Science Studies*, *6*, 1002–1038. https://doi.org/10.1162/QSS.a.16
Kwiek, M., & Szymula, L. (2026). Women in science: Measuring participation in Europe across disciplines, generations and over time. *Humanities and Social Sciences Communication*, *13*, 759, 1–16. https://doi.org/10.1057/s41599-026-06912-x
Larivière, V., Gingras, Y., Sugimoto, C. R., & Tsou, A. (2015). Team size matters: Collaboration and scientific impact since 1900. *Journal of the Association for Information Science and Technology*, *66*(7), 1323–1332.
Leydesdorff, L., & Bornmann, L. (2011). Integrated impact indicators compared with impact factors: An alternative research design with policy implications. *Journal of the American Society for Information Science and Technology*, *62*(11), 2133–2146.
Leydesdorff, L., Wagner, C. S., & Bornmann, L. (2019) Interdisciplinarity as diversity in citation patterns among journals: Rao-Stirling diversity, relative variety, and the Gini coefficient. *Journal of Informetrics*, *13*(1), 255–269.
Lindahl, J. (2018). Predicting research excellence at the individual level: The importance of publication rate, top journal publications, and top 10% publications in the case of early career mathematicians. *Journal of Informetrics*, *12*(2), 518–533.
Melguizo, T., & Strober, M. H. (2007). Faculty salaries and the maximization of prestige. *Research in Higher Education*, *48*(6), 633–668.
Merton, R. K. (1968). The Matthew effect in science. *Science*, *159*(3810), 56–63.
Moed, H. F. (2005). *Citation analysis in research evaluation*. Springer.
O'Boyle Jr., E., & Aguinis, H. (2012). The best and the rest: Revisiting the norm of normality of individual performance. *Personnel Psychology*, *65*(1), 79–119.
Parker, J. N., Lortie, C., & Allesina, S. (2010). Characterizing a scientific elite: The social characteristics of the most highly cited scientists in environmental science and ecology. *Scientometrics*, *85*(1), 129–143.
Petersen, A. M., Riccaboni, M., Stanley, H. E., & Pammolli, F. (2012). Persistence and uncertainty in the academic career. *Proceedings of the National Academy of Sciences*, *109*(14), 5213–5218.
Pinski, G., & Narin, F. (1976). Citation influence for journal aggregates of scientific publications: Theory, with application to the literature of physics. *Information Processing & Management*, *12*(5), 297–312.
Piro, F. N., Aksnes, D. W., & Rørstad, K. (2013). A macro analysis of productivity differences across fields: Challenges in the measurement of scientific publishing. *Journal of the American Society for Information Science and Technology*, *64*(2), 307–320.
Price, D. J. de Solla. (1963). *Little science, big science*. Columbia University Press.
Price, D. J. de Solla. (1976). A general theory of bibliometric and other cumulative advantage processes. *Journal of the American Society for Information Science*, *27*(5), 292–306.
Radicchi, F., Fortunato, S., & Castellano, C. (2008). Universality of citation distributions: Toward an objective measure of scientific impact. *Proceedings of the National Academy of Sciences*, *105*(45), 17268–17272.
Rosen, S. (1981). The economics of superstars. *The American Economic Review*, *71*(5), 846–858.
Seglen, P. O. (1992). The skewness of science. *Journal of the American Society for Information Science*, *43*(9), 628–638.
Serenko, A., Cox, R. A. K., Bontis, N., & Booker, L. D. (2011). The superstar phenomenon in the knowledge management and intellectual capital academic discipline. *Journal of Informetrics*, *5*(3), 333–345.
Shibayama, S., & Baba, Y. (2015). Impact-oriented science policies and scientific publication practices: The case of life sciences in Japan. *Research Policy*, 44(4), 936–950.
Sidiropoulos, A., Gogoglou, A., Katsaros, D., & Manolopoulos, Y. (2016). Gazing at the skyline for star scientists. *Journal of Informetrics*, *10*(3), 789–813.
Stallings, J., Vance, E., Yang, J., Vannier, M. W., Liang, J., Pang, L., Dai, L., Ye, I., & Wang, G. (2013). Determining scientific impact using a collaboration index. *Proceedings of the National Academy of Sciences*, *110*(24), 9680–9685.
Stephan, P. (2012). *How economics shapes science*. Harvard University Press.
Todeschini, R., & Baccini, A. (2016). *Handbook of bibliometric indicators: Quantitative tools for studying and evaluating research.* Wiley.
Wagner, C. S., Park, H. W., & Leydesdorff, L. (2015). The continuing growth of global cooperation networks in research: A conundrum for national governments. *PLOS ONE, 10*(7), e0131816.
Waltman, L., & van Eck, N. J. (2013). A systematic empirical comparison of different approaches for normalizing citation impact indicators. *Journal of Informetrics*, *7*(4), 833–849.
Waltman, L. (2016). A review of the literature on citation impact indicators. *Journal of Informetrics, 10*(2), 365–391.
Wildgaard, L. (2016). A critical cluster analysis of 44 indicators of author-level performance. *Journal of Informetrics*, *10*(4), 1055–1078.
Wildgaard, l. (2019). An overview of author-level indicators of research performance. In W. Glänzel, H. F. Moed, U. Schmoch, & M. Thelwall (Eds.), *Springer handbook of science and technology indicators* (pp. 361–396). Springer.
Wildgaard, L., Schneider, J. W., & Larsen, B. (2014). A review of the characteristics of 108 author-level bibliometric indicators. *Scientometrics*, *101*(1), 125–158.
Whitley, R. (2000). *The intellectual and social organization of the sciences*. Oxford University Press.
Wuchty, S., Jones, B. F., & Uzzi, B. (2007). The increasing dominance of teams in production of knowledge. *Science, 316,* 1036–1039.
Xie, Y. (2014). 'Undemocracy': Inequalities in science. *Science*, *344*(6186), 809–810.

**SHORT BIO:**

Marek Kwiek, full professor and chairholder, UNESCO Chair in Institutional Research and Higher Education Policy, University of Poznan, Poland. A Visiting Researcher at the German Center for Higher Education Research and Science Studies (DZHW) in Berlin in 2022-2024. He spent three years at North American universities. His research area is quantitative studies of science and sociology of science; and his current focus is on women in science, social stratification in science, and global academic elites. He collected 2.7 million EUR for his research projects. His recent articles are in "Higher Education", "Studies in Higher Education", "Quantitative Science Studies", "Journal of Informetrics" and "Scientometrics". His work on attrition and retention in global science was featured in "Nature News" (2024, 2025) and his article on research productivity classes in "Nature" and "Times Higher Education" (2026). An international science policy expert for the European Commission, USAID, OECD, World Bank, UNESCO, Council of Europe, and numerous national governments. An ordinary member of the European Academy of Sciences and Arts and *Academia Europaea*. He is listed among the top 2% of most highly cited researchers in all seven editions (2019-2026).

# Electronic Supplementary Material: Nonlinear Journal Prestige Normalized Productivity: Defining and Measuring It

Marek Kwiek
Center for Public Policy Studies, Adam Mickiewicz University, Poznan, Poland
kwiekm@amu.edu.pl, ORCID: orcid.org/0000-0001-7953-1063

## 1. Sensitivity Analysis of the Parameter *k*[1]

Using publication, data I present JPNP calculations for a range of values of k and provide comparisons of hypothetical success scenarios. The interpretations of both the empirically based sensitivity analysis and the scenario-based outcomes will be done separately . This analysis utilizes actual publication history by Polish scientists that have been employed to analyze previous publications by the authors (Kwiek and Roszka (2026): "Top Performers and Top Journals: Persistent Concentration in Scientific Publishing", a preprint available at www.arXiv.org).

## 2. Data and Construction of JPNP(*k*) Variants

JPNP was calculated on the author-period-ASJC (subject area) level for the values k = {1.0, 1.1,...,4.0}, compared to k=2.5 as a base. For every period-subject area cell, I have been computing the ranks, percentiles, deciles, and memberships into the top 10%, top 5%, and top 1% elite group of researchers by productivity. The above calculations were made using a full counting productivity version of JPNP $C_{ij} = 1$. Therefore, each article has an author's results weighted with full prestige $x_j^k$, and is therefore unweighted in terms of co-authorship. The conclusions regarding the stability relate to the complete JPNP variant and do not test a fractionally weighted JPNP.

**Table 1. Characteristics of the data used in the sensitivity analysis of parameter *k***

| Dimension / statistic | Value | Comment |
|---|---|---|
| Source object | pubs_long_cs_full | The most complete publication object from the Top Performers and Top Journals project |
| Number of rows | 1,263,759 | Author–publication rows |
| Number of authors | 144,314 | Unique author identifiers |
| Number of publications | 433,546 | Unique publication EID identifiers |
| Number of periods | 5 | Publication periods available in the object |
| Number of ASJC groups | 15 | Fields / disciplinary groups |
| Minimum percentile | 0 | Lower bound of prestige |

[1] This Electronic Supplementary Material was prepared by Dr. Wojciech Roszka from the Poznan CPPS Team using the dataset used in Kwiek & Roszka (2026), for which I am very grateful.

| | | |
|---|---|---|
| Lower quartile of percentile | 46 | 25% of observations have a percentile no higher than this value |
| Median percentile | 70 | Half of observations have a percentile no higher than this value |
| Maximum percentile | 99 | Upper bound of prestige |

For each individual publication, the journal prestige percentile was rescaled to the interval $[0, 1]$. Then, for author $i$ in the period $t$, the following variant of the JPNP indicator was calculated:

$$JPNP_i(k) = \sum_j p_{ij}^k ,$$

where $p_{ij}$ denotes normalized journal prestige percentile for publication $j$ by author $i$. The parameter $k$ regulates how strongly publications located in journals located in the upper echelons of the prestige distribution are rewarded.

**Table 2. Scope of analyses conducted for different values of parameter $k$**

| Element of analysis | Operationalization | Purpose |
|---|---|---|
| Ranking stability | Spearman rank correlation relative to $k = 2.5$ | To assess whether changing $k$ leads to changes in the hierarchy of researchers |
| Stability of top performers membership | Jaccard index for the top 10%, top 5%, and top 1% of scientists in terms of productivity | To assess the stability of the composition of research productivity elite at selected levels |
| Entries into and exits from the top 10% | Share of researchers losing top 10% status | To assess the practical scale of changes in classification |
| JPNP distribution | Median, mean, IQR, and Q90 | To assess the impact of $k$ on the scale and concentration of the JPNP indicator |
| Profile components | Correlations between JPNP($k$) and measures of volume and prestige | To assess the shift from publication numbers (volume) to journal prestige |
| Latent success | Correlations with three success scenarios | To examine how JPNP behaves when different definitions of success are applied |

### 3. Ranking Stability Relative to $k = 2.5$

To assess the stability of rankings of scientists, I calculated Spearman's rank correlation coefficient (rho), which shows how closely the rankings of the JPNP(k) are to the reference rankings produced when k = 2.5 I applied. The rho of these two sets of scientists' rankings are very high for both extreme values of k (k= 1 and k= 4). High rho means that there is a similar ordering of researchers in terms of their performance in the two lists. It does not mean that researchers have maintained the same position, though. In addition, a researcher may cross over from being part of the upper echelon to being part of lower levels of performance. To provide additional information on the ranking stability, I also assessed the degree to which the groups formed under each k were consistent with one another. This gave me some idea as to whether the parameters produce a different set of researchers, rather than simply producing slightly different rankings of the same researchers.

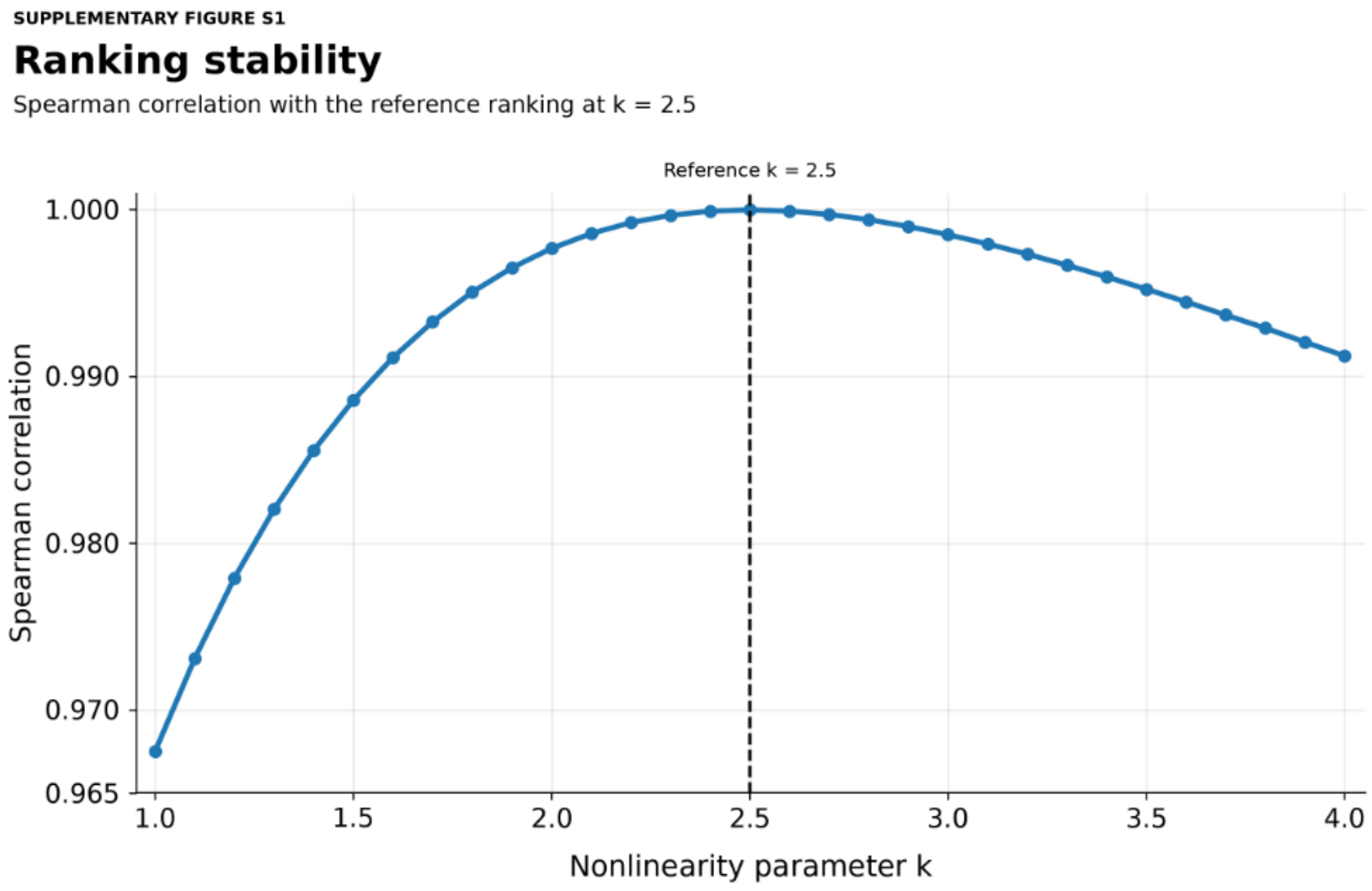

**Figure 1. Stability of the JPNP ranking with respect to the choice of parameter *k***

## 4. Stability of Top-Performer Composition

In the next stage, the composition of top performer groups were compared. The Jaccard Index was calculated between each group of top performers (the first 10%, then the first 5% and finally the first 1%) and the corresponding reference group at k = 2.5. The smooth behavior of these three indices when moving away from k = 2.5 shows that there is no break in the behavior of the top-performing groups.

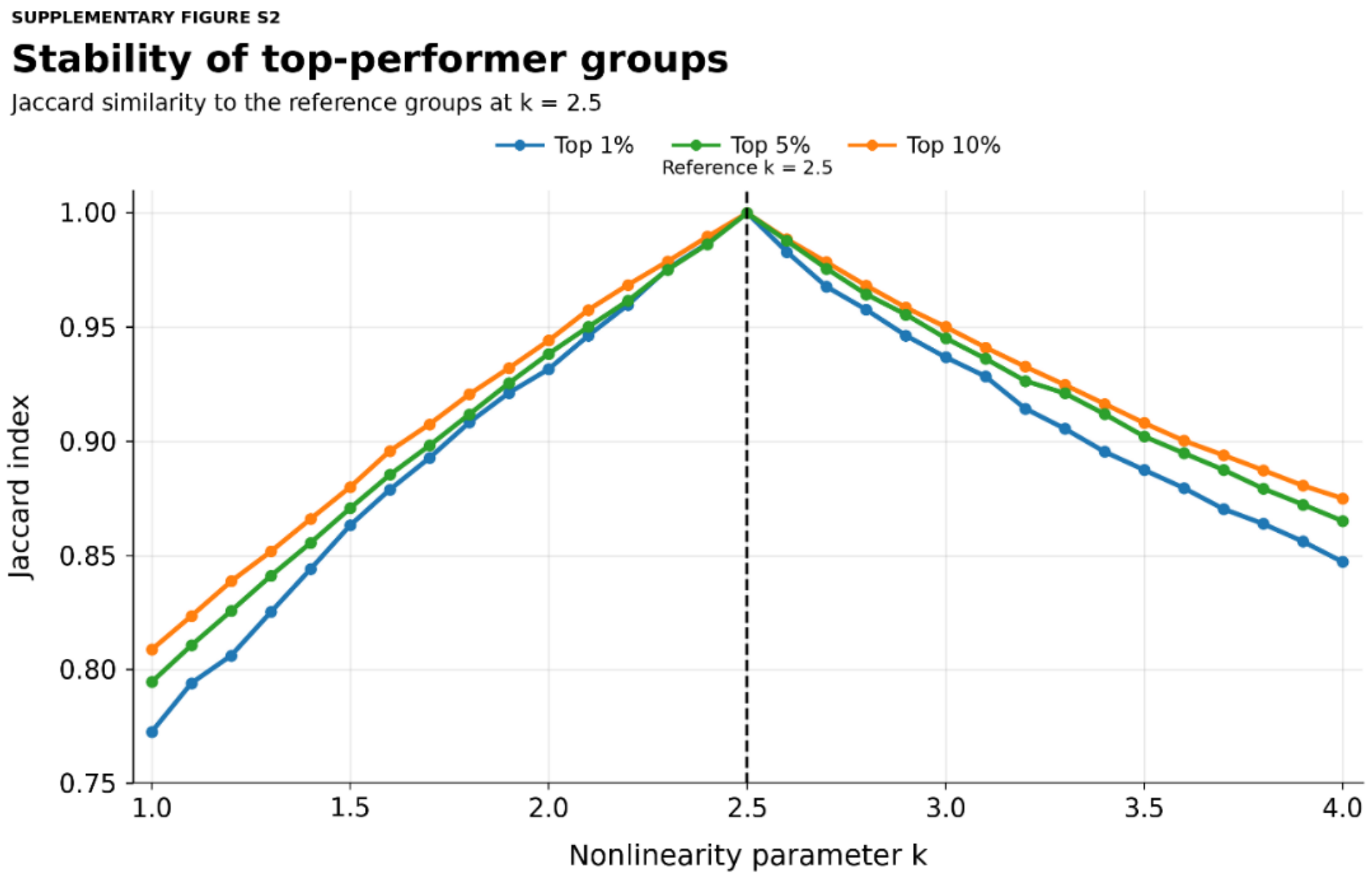

**Figure 2. Stability of top-performer composition with respect to the choice of parameter *k***

**Source for all Figures:** Own calculations based on the Observatory of Polish Science dataset used in Kwiek & Roszka (2026), "Top Performers and Top Journals: Persistent Concentration in Scientific Publishing," preprint available at: https://arxiv.org/abs/2603.00069.

## 5. The change of membership to the top 10% and changing k

Entries and exists from the top 10% class mean the change of classification of the same scientists after the application of a different k to the same data. These are not transitions between career periods or entries into science and leaving it. Beginning with a focus on the practical scale of the changes in classification, I analyzed how many researchers that are among the top 10% for k = 2.5 fall out of that position when an alternative value of k is applied. For k approaching values that would be considered "more linear", the number of researchers falling from being one of the top 10% increased at a faster rate than did when k was moving above 2.5.

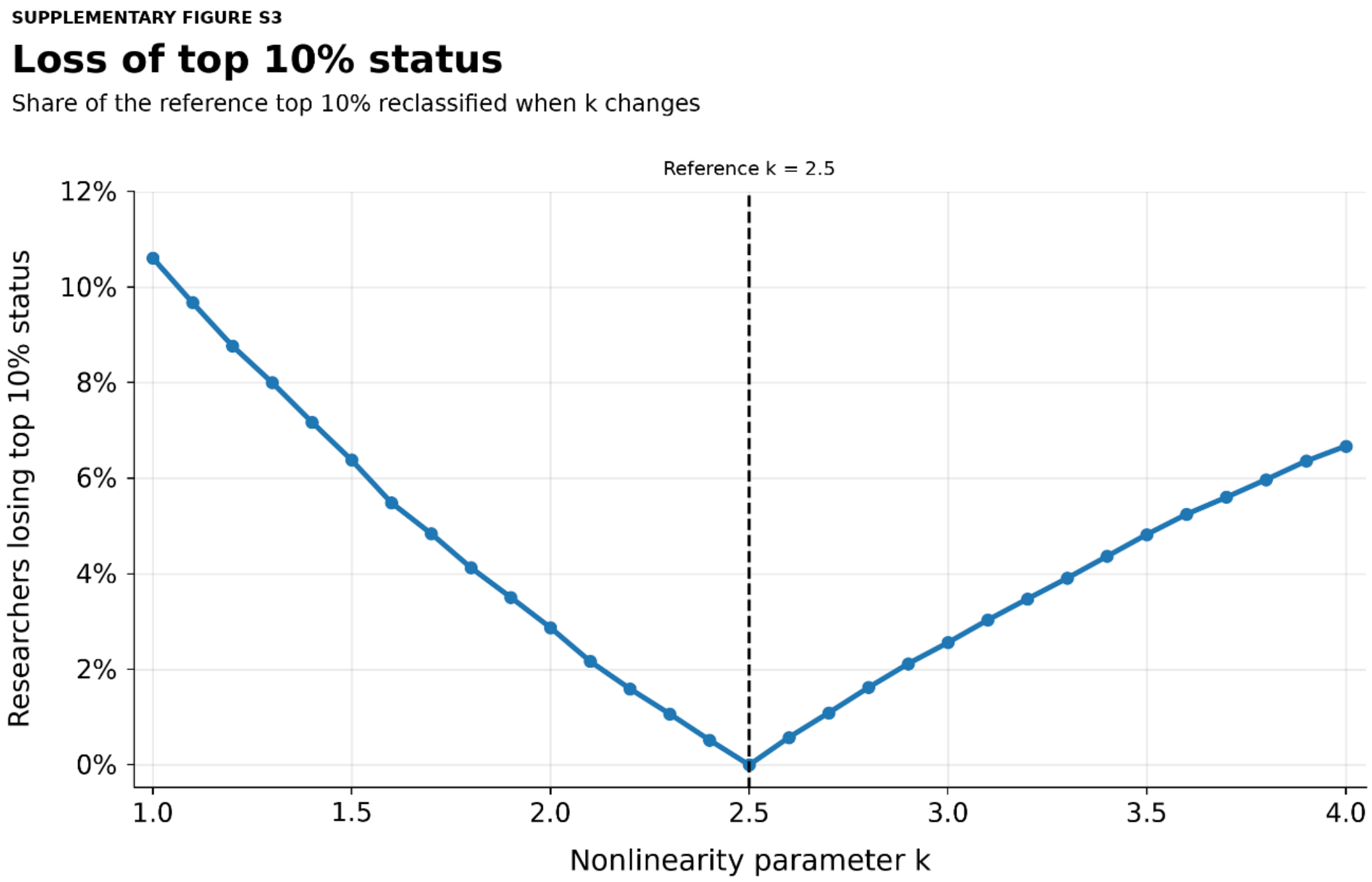

**Figure 3**. Change in Status Relative to Classification for k = 2.5 (Top 10%)

## 6. Distribution of JPNP By Value of k

This section examines how varying k will affect the degree and strength at which of JPNP. The variation in k has no effect on any the value of the individual JPNP. Therefore, the decrease in the mean and median values are a direct result of the way the function is constructed. It is not a proof of the increase in concentration or a higher accuracy of the indicator.

For better clarity in reading, therefore, instead of using multi-plot density plots for each k level, I have shown here the distributions of basic statistical properties: the median, mean, inter-quartile range and the 90th percentile.

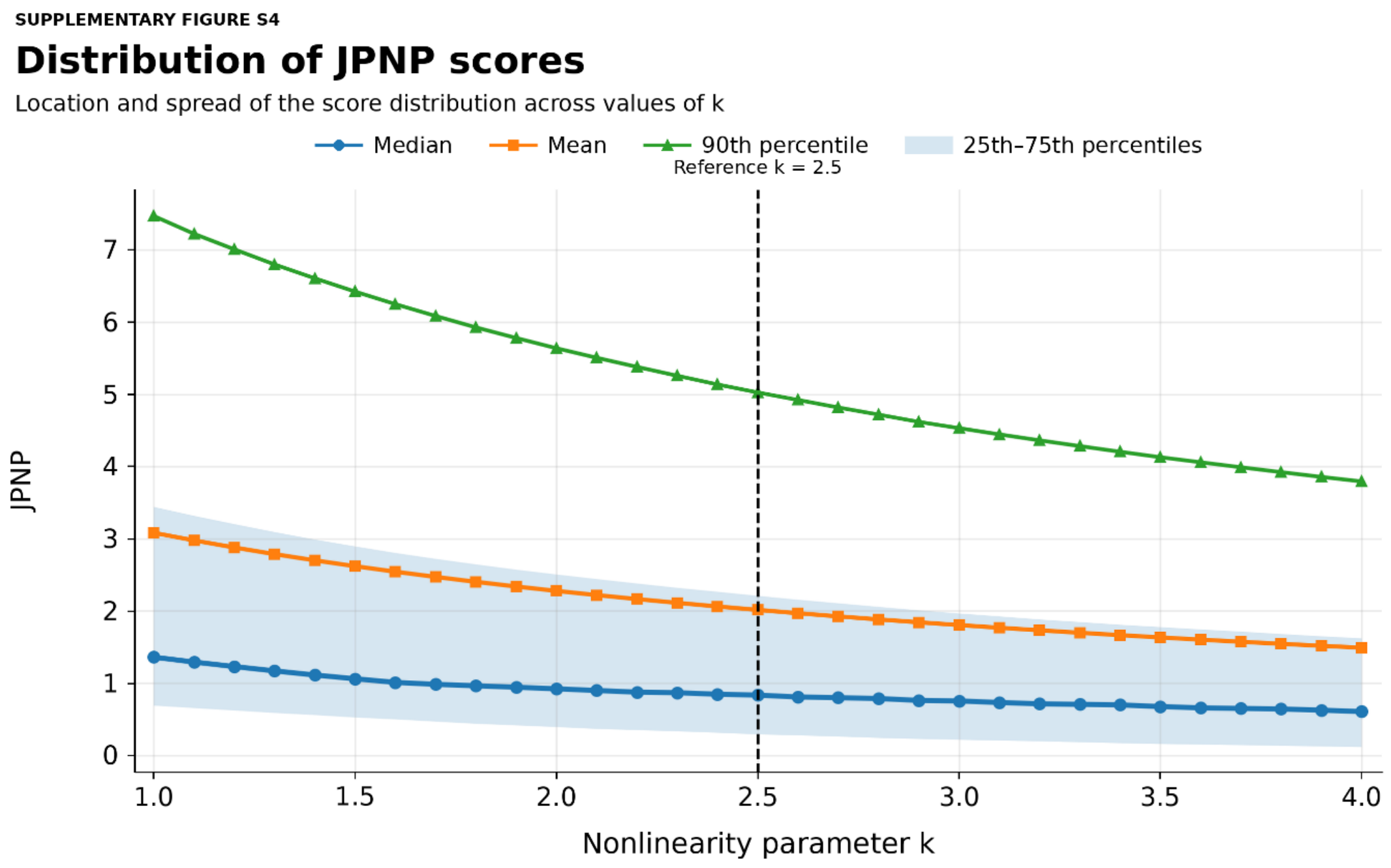

**Figure 4. JPNP distribution across values of parameter $k$**

## 7. Agreement of JPNP($k$) with Publication-Profile Components

In the results presented, as k increases, the correlation of JPNP and the number of publications decreases, and correlations with the prestige measures of the portfolio measured increase. It is a characteristics of the dataset examined rather than a guaranteed result for each population.

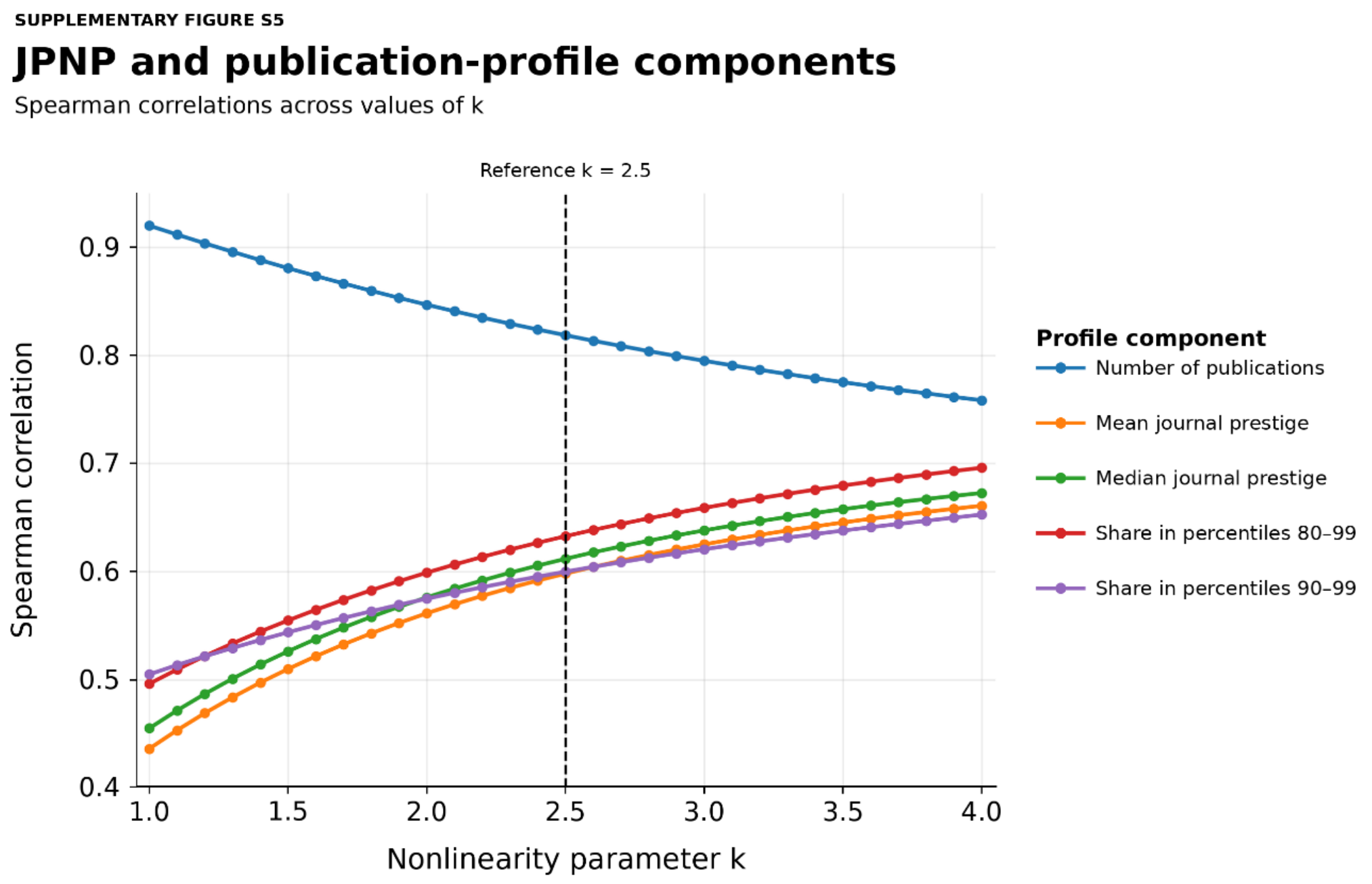

**Figure 5. Conformity of JPNP($k$) with publication-profile components**

# 8. Simulation of Latent Success

I analyze here whether there is an alignment between JPNP(k), and three hypothetical models for assessing latent scientific success. In the volume model, success was based on the total amount of publications. In the balanced model, it included a combination of total number of publications and the percentage of publications among top percentiles. And finally, in the prestige model, the most important factor was the presence of publications in highly prestigious journals. Scenarios are hypothetical criteria constructed using the components of publication profiles. The conformity of JPNP with these criteria shows the behavior of the indicator under different assumptions – but it is not a validation with respect to independently observed scientific success.

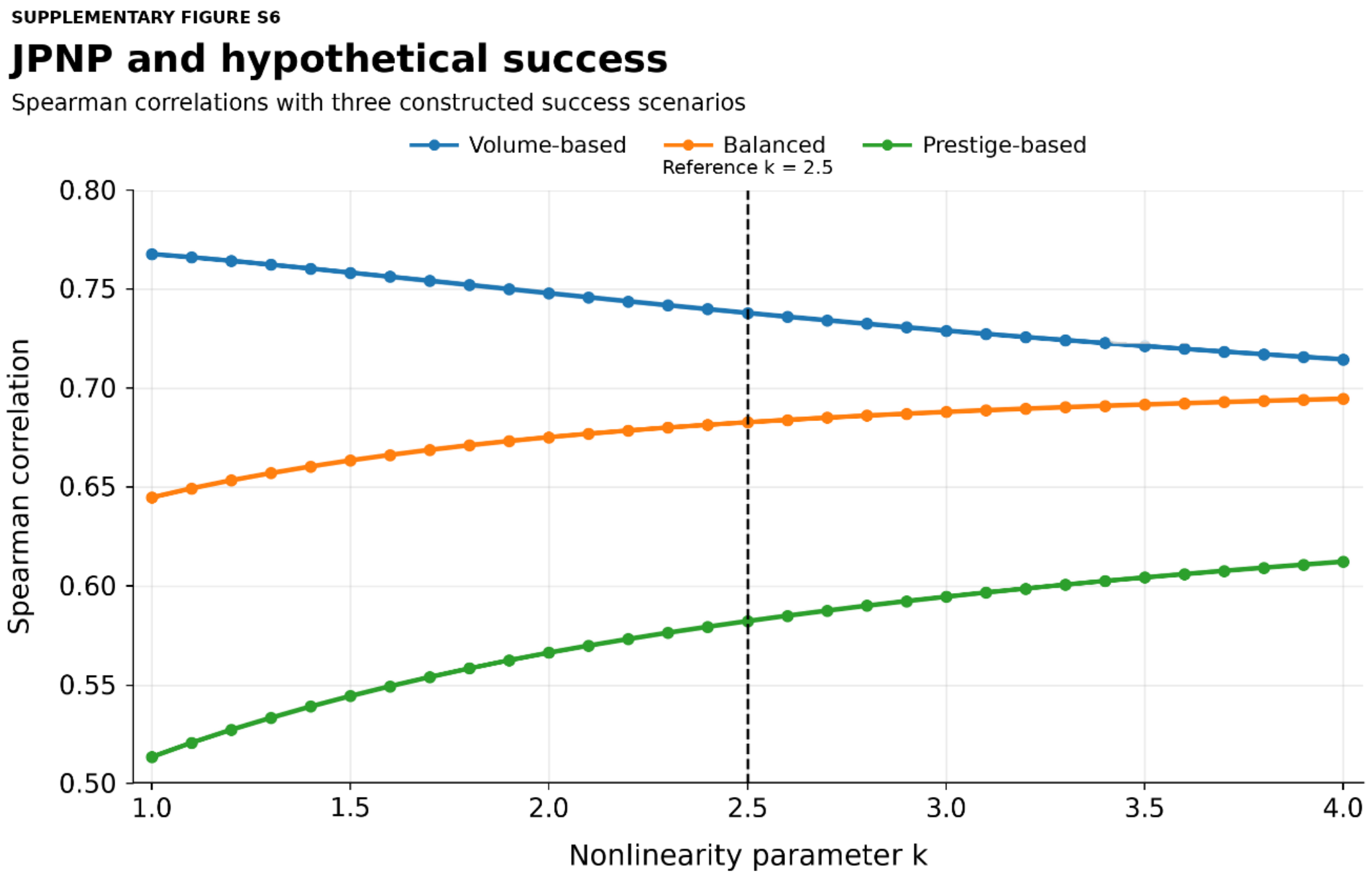

**Figure 6. Conformity of JPNP($k$) with latent scientific success**

**Table 3. Direction of the relationships between the parameter $k$ and components of the publication profile**

| **Profile component** | **Change in correlation as $k$ increases** | **Interpretation** |
|---|---|---|
| Number of publications | Decreases | Higher $k$ weakens the role played by publication volume |
| Journal prestige - mean | Increases | Higher $k$ strengthens the role of the average journal quality |
| Journal prestige - median | Increases | Higher $k$ rewards a consistently high-prestige individual publication profile |
| Share of publications in 80–99 percentiles | Increases | Higher $k$ rewards the presence of publications in the upper echelons of the prestige distribution |
| Share of publications in 90–99 percentiles | Increases | Higher $k$ increases selectivity toward the highest-prestige journals |

**Table 4. Stable ranges and summary justification for choosing $k = 2.5$**

| **Criterion** | **Result of the analysis** | **Implication for JPNP** |
|---|---|---|
| Stable Spearman range, tolerance 0.005 | 1.8–3.5 | A broad range of values is practically equivalent in ranking terms |
| Stable Spearman range, tolerance 0.010 | 1.6–4.0 | Almost the entire analyzed range gives a ranking very close to the reference ranking |
| Stability of top performers | Smooth change, no breaks | Classification is robust to moderate changes in the parameter |
| Entries into/exits from the top 10% | Largest changes occur near $k = 1$ | Linear percentile weighting leads to a change in the composition of the top 10% elite relative to the k = 2.5 variant |
| Profile components | Shift from volume to prestige | $k = 2.5$ is a compromise between publication volume and outlet prestige |

**Source:** Own calculations, the Observatory of Polish Science dataset. We used the data in Kwiek & Roszka (2026), "Top Performers and Top Journals: Persistent Concentration in Scientific Publishing," preprint available at: https://arxiv.org/abs/2603.00069.

## 9. Summary of Findings and Justification for $k = 2.5$

My analysis shows that the presentation of k = 2.50 as a unique value (or as an obligatory mathematical value) is inappropriate. It should be presented as an important value among several values residing in a relatively stable parameter space. There remains a very strong correlation between researcher's ranking across a broad spectrum of k, and while there are some gradual changes with respect to group membership of the top performers, there are no sharp discontinuities observed. The k = 2.5 variant differs from the k = 1 variant that uses linear prestige weighting. It is not a comparison with simple publication counting in which each publication receives a prestige weight equal 1. In the dataset examined, k approaching 1 is linked with higher changes of the composition of the reference group of top 10%. The analysis supports the application of k = 2.5 as a reference value for the dataset studied and the JPNP variant used. It does not show that it is the optimal value or a value equally appropriate for all fields, countries, institutions and authorship variants.

### Reference

Kwiek, M., & Roszka, W. (2026). Top performers and top journals: Persistent concentration in scientific publishing. arXiv. https://arxiv.org/abs/2603.00069